\documentclass[12pt,preprint]{aastex}

\usepackage{amsmath}
\usepackage{color} 
\usepackage{ulem}
\usepackage{natbib}
 \newcommand{\co}{${\rm ^{12}CO}$~} 	
 \newcommand{\hco}{${\rm HCO^{+}}$\,} 	
 \newcommand{\kms}{km s$^{-1}\,$} 
 \newcommand{\3}{($J=3-2$)} 
 \newcommand{\4}{($J=4-3$)} 
 \newcommand{\Hii}{H{$\,${\sc ii}}~} 
  \newcommand{\Vlsr}{$V_{\rm LSR}$} 

\slugcomment{}

\shorttitle{W40}
\shortauthors{Shimoikura et al.}

\begin{document}


\title{Dense Clumps and Candidates for Molecular Outflows in W40}


\author{T{\scriptsize OMOMI} S{\scriptsize HIMOIKURA}\altaffilmark{1},
K{\scriptsize AZUHITO} D{\scriptsize OBASHI}\altaffilmark{1},
F{\scriptsize UMITAKA} N{\scriptsize AKAMURA}\altaffilmark{2,3}}

\author{C{\scriptsize HIHOMI} H{\scriptsize ARA}\altaffilmark{2,4}, 
T{\scriptsize OMOHIRO} T{\scriptsize ANAKA}\altaffilmark{5},
Y{\scriptsize OSHITO} S{\scriptsize HIMAJIRI}\altaffilmark{6}, 
K{\scriptsize OUJI} S{\scriptsize UGITANI}\altaffilmark{7}, 
\and 
R{\scriptsize YOHEI} K{\scriptsize AWABE}\altaffilmark{2,4,8}}
\affil{\scriptsize{\rm $^1$ ikura@u-gakugei.ac.jp}}

\altaffiltext{1}{Department of Astronomy and Earth Sciences, Tokyo Gakugei University, 
Koganei, Tokyo 184-8501, Japan}
\altaffiltext{2}{National Astronomical Observatory of Japan, Mitaka, Tokyo 181-8588, Japan}
\altaffiltext{3}{Nobeyama Radio Observatory, Minamimaki, Minamisaku, Nagano 384-1305, Japan}
\altaffiltext{4}{The University of Tokyo, 7-3-1 Hongo Bunkyo, Tokyo 113-0033, Japan}
\altaffiltext{5}{Department of Physical Science, Osaka Prefecture University, Sakai, Osaka 599-8531, Japan}
\altaffiltext{6}{Laboratoire AIM, CEA/DSM-CNRS-Universit${\rm \acute{e}}$ Paris Diderot, IRFU/Service d'Astrophysique, CEA Saclay, F-91191 Gif-sur-Yvette, France}
\altaffiltext{7}{Graduate School of Natural Sciences, Nagoya City University, Mizuho-ku, Nagoya 467-8501, Japan}
\altaffiltext{8}{SOKENDAI (The Graduate University for Advanced Studies), Shonan Village, Hayama, Kanagawa 240-0193, Japan}


\begin{abstract}

We report results of the \co\3 and \hco\4 observations of the W40 \Hii region
with the ASTE 10 m telescope (HPBW$\simeq22\arcsec$) 
to search for molecular outflows and dense clumps.
We found that the velocity field in the region is highly complex,
consisting of at least four distinct velocity components at $V_{\rm LSR} \simeq 3$, 5, 7, and 10 \kms.
The $\sim$7 \kms component represents the systemic velocity of cold gas surrounding
the entire region, and causes heavy absorption in the \co spectra
over the velocity range $6 \lesssim V_{\rm LSR} \lesssim 9$ \kms.
The $\sim$5 and $\sim$10 \kms components exhibit high \co temperature ($\gtrsim40$ K)
and are found mostly around the \Hii region, suggesting that these components
are likely to
be tracing dense gas interacting with the expanding shell around the \Hii region.
Based on the \co data, 
we identified 13 regions of high velocity gas which we interpret as candidate outflow lobes. 
Using the \hco data, we also identified six clumps and estimated their physical parameters.
On the basis of the ASTE data and near-infrared images from 2MASS, 
we present an updated three-dimensional model of this region. 
In order to investigate molecular outflows in W40,
the SiO ($J=1-0$, $v=0$) emission line and some other emission lines at 40 GHz 
were also observed with the 45 m telescope at the Nobeyama Radio Observatory,
but they were not detected at the present sensitivity.
\end{abstract}


\section{INTRODUCTION}
The W40 complex is a site of active star formation in the Aquila Rift
containing molecular clouds, star cluster, massive stars, and an \Hii region. 
Several mid and far-infrared observations have revealed that the 
W40 complex consists of a number of filamentary structures to form an hourglass shape 
\citep[e.g.,][]{Rodney, Kuhn, Andre,Bontemps,Konyves}.
There are a number of young stellar objects (YSOs) associated with this complex \citep[e.g.,][]{Rod,Kuhn, Maury, Mallick}. 
The cluster and its associated \Hii region are located to the northwest of the constriction of the hourglass. 
The cluster is found to have a considerably large stellar surface density of $\sim 650$ pc$^{-2}$ \citep{Mallick}. 
Several ionizing sources for the \Hii region are also found in this region. 
IRS 1A South is an O9.5 spectral type star \citep{Shuping} and is the most luminous source. 
Another high mass star, IRS 5, is located in the northwest of IRS 1A South. 
This is classified as a B1V type star \citep{Shuping, Mallick}. 
Radio emission from a roughly spherical compact \Hii region 
has been detected around IRS 5 by \cite{Mallick}.
They suggested that high-mass star formation is going on therein.

Previous molecular line and radio continuum observations 
revealed the presence of the dense gas around the W40 \Hii region \citep[e.g.,][]{Zeilik,Zhu, Pirogov, Maury}.
\cite{Dobashi2005} and \cite{Dobashi2011} named the dense region as TGU 279-P7 and No.1195, 
based on the Digitized Sky Survey (DSS) and the Two Micron All Sky Survey (2MASS) extinction data, respectively.
The dense region has a ring-like shape along the boundary of the \Hii region \citep{Maury}.
\cite{Pirogov} carried out multiple-molecular line observations 
such as HCO$^{+}$, CS, and NH$_3$, 
and found the large variations of the fractional abundances of these molecules over the dense region.
All of these studies indicate that the \Hii region
is strongly interacting with the surrounding dense gas. 
The W40 complex is therefore an excellent region to study 
the interaction of massive stars with their parental molecular cloud. 

Several studies have discussed the geometry of the W40 complex \citep[e.g.,][]{Crutcher,Vallee,Kuhn,Pirogov}.  
Especially, \cite{Vallee} suggested the blister model for W40 which is illustrated in their Figure 2 and
is now regarded as a standard picture of the region.
However, the complex velocity structure has prevented us from complete understanding 
of the geometry and kinematics of the molecular cloud associated with W40 \citep[e.g.,][]{Crutcher, Pirogov}.
More detailed studies to uncover the internal density distribution and kinematics of 
the molecular cloud is needed to better understand the structure of this region.


Low-mass members of this region were also revealed by \cite{Maury} and \cite{Mallick}. 
They identified a number of Class 0/I sources and starless cores around the \Hii region. 
However, outflow activity in this region remains uncertain.  
\cite{Zhu} discovered a weak molecular outflow 
through their \co\3 observations with the KOSMA 3 m telescope 
whose half power beam width (HPBW) is $\sim80\arcsec$. 
The identified outflow lobes appear to have complicated structure. 
In addition, the surveyed area is limited to a small region just around the \Hii region. 
It is important to carry out a comprehensive survey for molecular outflows covering much wider area 
to reveal the star formation activity in this region.

In this paper, 
we present results of the \co\3 and \hco\4 observations made 
toward a large area ($\sim8\arcmin \times 8\arcmin$) covering the whole W40 \Hii region 
and its interacting molecular cloud, 
using Atacama Submillimeter Telescope Experiment (ASTE) 10 m telescope. 
For the purpose of an outflow survey, we carried out \co\3 mapping observations to
detect 13 outflow candidates. 
In order to explore the velocity and spatial distributions of the dense gas in the W40 region,
we also carried out \hco\4 observations to identify six clumps, and further attempted to model their
locations relative to the \Hii region along the line-of-sight mainly through a comparison with
the 2MASS near-infrared images.

Hereafter we will refer to the observed region simply as W40.
The distance estimation to W40 remains uncertain, 
and varies from 340 to 600 pc depending on the methods used \citep[e.g.,][]{Rodney,Kuhn, Shuping}.
Here, we adopt the latest estimate of 500 pc derived by \cite{Shuping} throughout this paper.


\section{OBSERVATIONS} \label{obs}

\subsection{$^{12}$CO and HCO$^{+}$ Observations with the ASTE 10 m Telescope} \label{ASTEobs}

The mapping observations of W40 were performed with the \co($J=3-2$) and \hco($J=4-3$) emission
lines at 345.795991 GHz and 356.734242 GHz, respectively,
using the ASTE 10 m telescope \citep{Ezawa,Kohno} at Atacama, Chile. 
Data of these molecular lines were obtained simultaneously 
during the period between 2013 September 26th and October 1st.
The area observed is denoted in Figure \ref{fig:iimap}, which covers the central part of the W40 \Hii region. 
The angular resolution of the telescope is HPBW$\simeq22\arcsec$ at the observed frequencies.
A side-band separating (2SB) mixer receiver CATS345 was used for the observations.
The spectrometer was an XF-type digital spectro-correlator, called MAC, with 1024 channels 
covering a bandwidth of 128 MHz. 
The frequency resolution was 125 kHz, corresponding to a velocity resolution
of 0.1 \kms at the observed frequencies.
The system noise temperature $T_{\rm sys}$ was $300-400$ K during the observations.
The observations were made using the on-the-fly (OTF) mapping technique \citep{Sawada}.
Data were calibrated with the standard chopper-wheel method \citep{Kutner}. 
The telescope pointing was regularly checked every 1.5 hour using \co\3 line from W-Aql, 
and the pointing error was smaller than $2\arcsec$ during the observations. 
In addition, we checked the intensity reproducibility by observing
M17SW {[}R.A. (2000) = 18$^{h}$20$^{m}$23$^{s}$.08, Dec. (2000) =$-16{\arcdeg}$11${\arcmin}$43.3${\arcsec}${]} 
\citep{Wang1994}
and  
Serpens South {[}R.A. (2000) = 18$^{h}$30$^{m}$03$^{s}$.8, Dec. (2000) =$-02{\arcdeg}$03${\arcmin}$03.9${\arcsec}${]} 
\citep{Nakamura2011}
every day to calibrate the spectral data, which fluctuates within $\sim10\%$.

Data reduction and map generation were performed using the software package NOSTAR
developed by the Nobeyama Radio Observatory (NRO). 
The data were convolved with a spheroidal function 
\footnote{\cite{Schwab} and \cite{Sawada} described the details of the spheroidal function. 
We applied the parameters, $m = 6$ and $\alpha = 1.0$, which define the shape of the function.} 
and resampled onto a uniform grid of $10\arcsec$,
and then corrected for the main-beam efficiency of the telescope ($\eta_{\rm mb}=0.6$).
We finally obtained the spectral data in the brightness temperature scale $T_{\rm mb}$ in Kelvin 
with an effective angular resolution of approximately $31\arcsec$ 
and a velocity resolution of 0.1 \kms for each emission line. 
Typical noise level of the spectra was $\Delta T_{\rm mb} =0.2$ K for this velocity resolution. 

\subsection{SiO, CCS, HC$_{3}$N, and HC$_{5}$N Observations with the NRO 45 m Telescope} \label{45obs}
We also carried out mapping observations toward W40 in the SiO ($J=1-0$, $v=0$) 
emission line at 43.42376 GHz using the NRO 45 m telescope on 2014 May 27th and 28th. 
The SiO molecule is a well-known probe of molecular outflows \citep[e.g.,][]{Hirano2006}. 
The molecule is also one of the shock tracers in molecular outflows 
which was first evidenced in L1157 by \citet{Mikami}, and has been widely used as the shock tracer for other star forming regions
\citep[e.g.,][]{Shimajiri2008,Shimajiri2009,Nakamura2014}.
In order to search for the possible SiO emission associated with the outflows in W40, 
we performed SiO mapping observations over the same regions observed with the ASTE telescope. 

We used a dual-polarization receiver Z45 \citep{Tokuda}
and the digital spectrometers SAM45 with 4096 channels covering a bandwidth of 125 MHz. 
Combination of the receiver and spectrometers made it possible to observe 
the SiO, CCS ($J_{N}$=4$_{3}-3_{2}$, 45.379033 GHz),
HC$_{3}$N ($J=5-4$, 45.490316 GHz), and HC$_{5}$N ($J=17-16$, 45.264721 GHz) emission lines simultaneously. 
The channel width of the spectrometers was 3.81 kHz, 
corresponding to a velocity resolution of $\sim 0.025$ \kms at 45 GHz. 
The observations were performed using the OTF mapping technique and 
the data were calibrated with the standard chopper-wheel method. 
$T_{\rm sys}$ was typically $180$ K, and the pointing accuracy was better than $10 \arcsec$ during the observations, 
as was checked by observing the SiO maser IRC+00363 at 43 GHz every 1.5 hours.
$\eta_{\rm mb}$ and HPBW were 0.7 and $37\arcsec$, respectively. 
Data reduction and analysis were done in the same manner as for the ASTE observations.
We finally obtained the spectral data with an effective angular resolution of approximately $57\arcsec$ 
and a velocity resolution of 0.05 \kms. 
The resulting $1\,\sigma$ noise level for this velocity resolution was $\Delta T_{\rm mb}\sim 0.2$ K, 
but no significant SiO and other molecular emission was detected at this sensitivity.


\section{RESULTS AND DISCUSSION}


\subsection{Distributions of Molecular Gas} \label{sec:gas}

\begin{figure*}
\begin{center}
\includegraphics[scale=0.75]{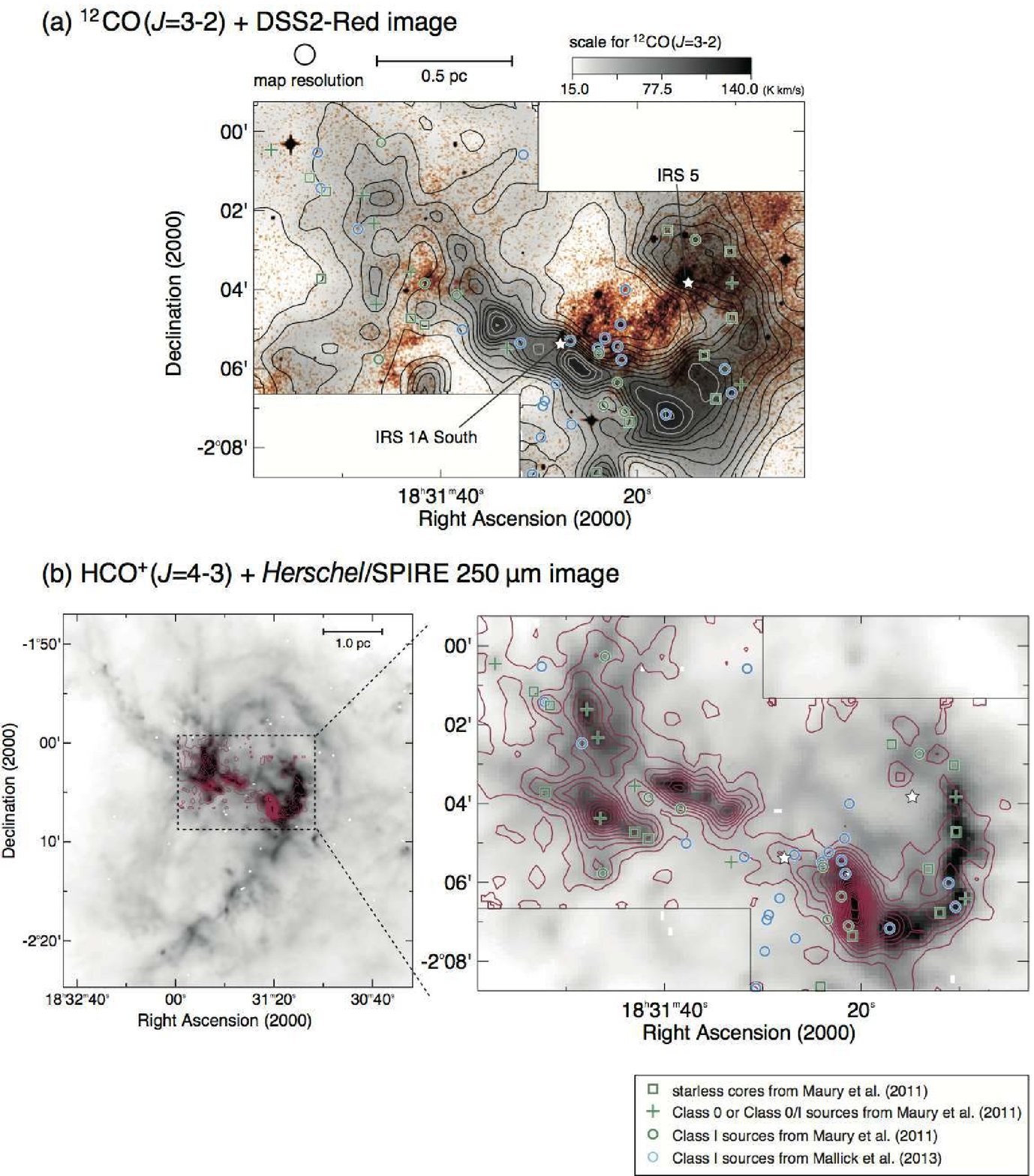}
\caption{
Integrated intensity maps of the (a) \co\3 emission line and (b) \hco\4 emission line around W40 obtained with the ASTE telescope. 
Velocity ranges used for the integration are $-3.0 \leqq V_{\rm LSR} \leqq 15.0$ \kms 
and $3.0 \leqq V_{\rm LSR} \leqq 11.4$ \kms for the \co and \hco maps, respectively. 
The lowest contours and the contour intervals for the map in panel (a) are 10 K \kms, and those in panel (b) are 1 K \kms.
The red color scale in panel (a) is the DSS2-Red image, and the gray scale in panel (b) is the {\it Herschel} SPIRE 250 $\micron$ image.
Positions of the high mass IRS sources \citep[IRS 1A Sourth and IRS 5,][]{Shuping} are indicated by star marks. 
Squares, plus signs, and circles denote the starless cores and Class 0/I sources classified by \cite{Maury} (colored in green) and \cite{Mallick} (colored in blue).
The map resolution ($31\arcsec$) and a linear scale of 0.5 pc (at $D$=500 pc) are shown at the top of panel (a). 
The area enclosed by the thin lines in panel (a) and right side of panel (b) denotes the observed region.
\label{fig:iimap}}
\end{center}
\end{figure*}

Figure \ref{fig:iimap} (a) shows the \co\3 intensity distribution integrated over the velocity range 
$-3.0\leqq V_{\rm LSR}\leqq15.0$ \kms superposed on the DSS2-Red image.
The intensity distribution in the south-west part of the map shows an arc-shaped structure which
can be identified also in other molecular maps in the literature \citep[e.g.,][]{Zhu, Pirogov}.
Our map shows that the \co emission is extending to the north-east direction,
which has not been revealed by molecular observations with a sub-arcmin angular resolution to date.
In the figure, we also show the positions of known YSOs, i.e., 
Class 0/I sources (plus signs or circles) and starless cores (squares) 
found by \cite{Maury} through the 1.2 mm dust continuum observations as well as Class I sources (circles) 
identified by \citet{Mallick} from the archival {\it Spitzer} observations in conjunction with near-infrared data from UKIRT.
The high mass sources IRS 1A South and IRS 5 are also shown with white stars in Figure \ref{fig:iimap} (a). 
It is clearly seen that the YSOs are distributed 
along the arc-shaped structure where the \co emission is enhanced up to $T_{\rm mb}(^{12}{\rm CO})=40-50$ K,
suggesting that the arc-shaped structure is the dense gas heated and/or compressed by the ionization front of the \Hii region.

Figure \ref{fig:iimap} (b) shows the \hco\4 intensity distribution integrated over the velocity range 
$3.0\leqq V_{\rm LSR}\leqq11.4$ \kms 
superposed on the {\it Herschel} SPIRE $250 \,\micron$ image 
\footnote{{\it Herschel} SPIRE image which was obtained as part of Proposal ID:SDP\_pandre\_3.} 
(colored in gray).
The \hco distribution is well correlated with the dust distribution traced by the {\it Herschel} data 
and 1.2 mm continuum image obtained by \citet[][see their Figure 3]{Maury}. 
It is noteworthy that the local peaks of the \hco map are coincident with or located 
close to massive dense clumps identified by \cite{Pirogov}
through the 1.2 mm continuum observations, although they mapped only a half of the area of our map. 
We found that the intense \hco emission is seen mainly in two parts in the map:
One is the arc-shaped structure in the south-west part, and the 
other is the filamentary structure in the north-east part. 
The central part of the cluster including IRS 1A South ionizing the \Hii region is located
in between the two parts. 
\color{black}
As seen in Figure \ref{fig:iimap} (b), YSOs are distributed around the dense gas traced in \hco, 
suggesting that the sources were formed therein.
The \hco map also shows that there is less dense gas around IRS 5, 
though several Class 0/I sources or starless cores have been detected there. 

We did not detect the CCS ($J_{N}$=4$_{3}-3_{2}$) emission in W40 at the present sensitivity.
The CCS molecule is known to be more abundant in an earlier stage of chemical evolution
\citep[e.g.,][]{Suzuki1992,Hirota2009,Shimoikura2012}.
It is interesting to note that the Serpens South cloud
located only $\sim20\arcmin$ to the west of W40 exhibits strong
CCS emission ($T_{\rm mb} \gtrsim2$ K) as reported by \citet{Nakamura2014}.
They estimated the column density and fractional
abundance of CCS to be $N$(CCS)$\simeq 1\times10^{14}$ cm$^{-2}$ and 
$f$(CCS)$=10^{-10} - 10^{-9}$ in the Serpens South cloud.
From the noise level of our CCS data, we estimate the upper limits of
$N$(CCS) and $f$(CCS) to be $\sim1.1 \times 10^{13}$ cm$^{-2}$
and $\sim10^{-9.25}$ at the intensity peak position of the HCO$^+$ emission in W40, respectively
(see Appendix \ref{sec:N(CCS)}).
These values infer that $N$(CCS) in W40 is likely to be an order of magnitude smaller
than in the Serpens South cloud, which can be
the cause of the non-detection of the emission line.
The upper limit of $f$(CCS) is comparable to the fractional abundance
in the Serpens South cloud, suggesting that W40 may be more evolved
(or at a similar stage at most) in terms of the chemical reaction time.

\begin{figure*}
\begin{center}
\includegraphics[scale=0.75]{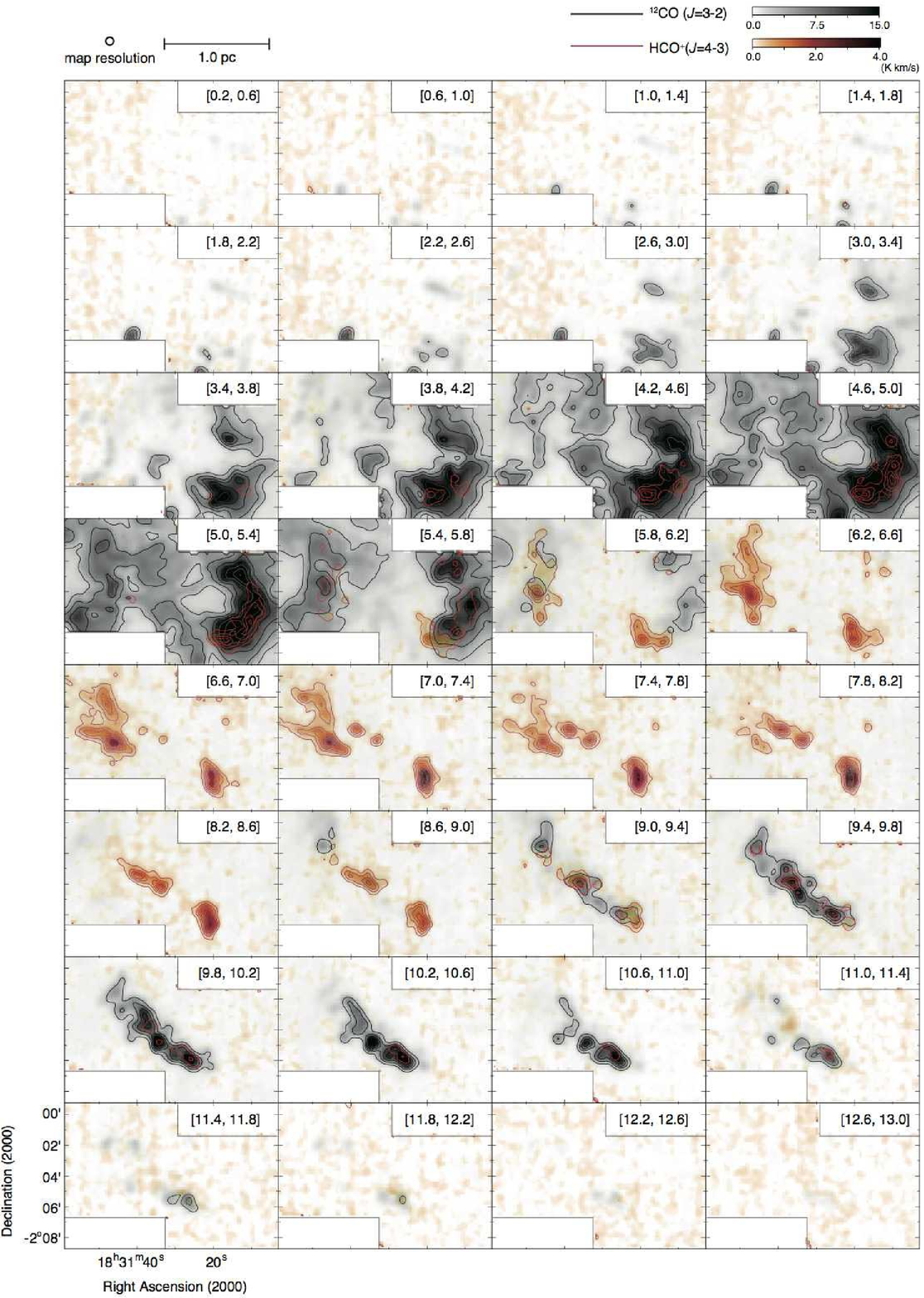}
\caption{
Channel maps of the \co\3 (gray) and \hco\4 (red) emission lines 
in the velocity range $0.2 \leqq V_{\rm LSR} \leqq 13.0$ \kms with the 0.4 \kms step.
The velocity range in units of \kms used for the integration is indicated at the top right corner of each panel. 
The lowest contours and the contour intervals are 3 K \kms for the \co maps, and are 0.3 K \kms for the \hco maps.
\label{fig:channel}}
\end{center}
\end{figure*}

\begin{figure*}
\begin{center}
\includegraphics[scale=1.]{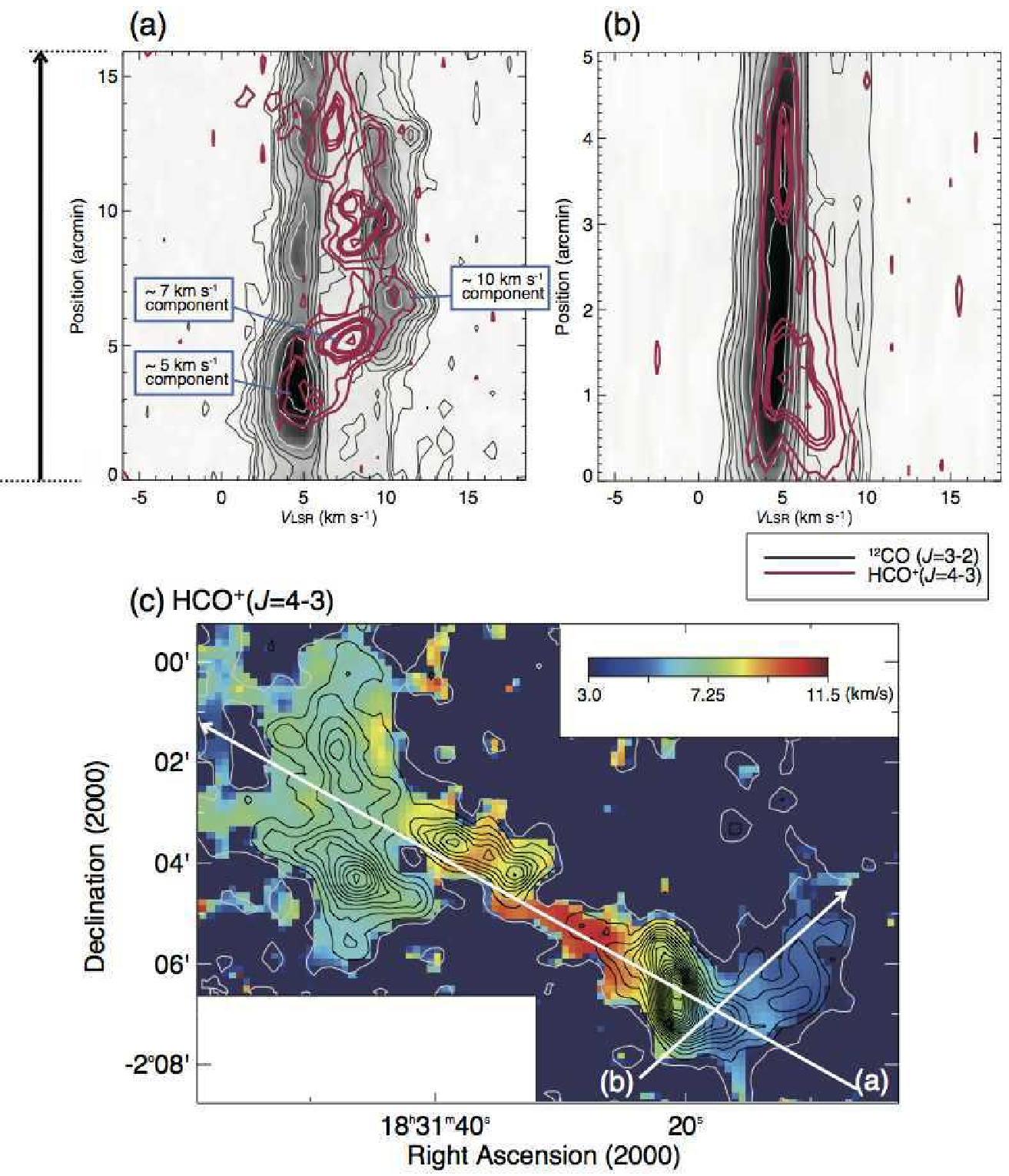}
\caption{
(a) Position-velocity diagrams of the \co\3 (black) and \hco\4 (red) emission lines taken along the cut labeled (a)
in panel (c).  
(b) Same as panel (a), but for the other cut in panel (c). Contours in the position-velocity diagrams
are drawn at 0.4, 0.8, 1, 2, 4, 8, 10, 20, 30, and 40 K for the \co emission line, and 
0.2, 0.4, 1.0, 1.2, 1.4, 2.0, 2.2, and 2.4 K for the \hco emission line. 
(c) Mean velocity map of the \hco emission line over the velocity range $3.0\leqq V_{\rm LSR}\leqq11.5$ \kms. 
Contours represent the \hco integrated intensity map same as in Figure \ref{fig:iimap}(b). 
\label{fig:V0}}
\end{center}
\end{figure*}

The velocity field of W40 shows a great complexity. 
Figure \ref{fig:channel} displays the velocity channel maps of the \co and \hco emission lines 
with a velocity interval of 0.4 \kms.
The \co emission is significantly detected in two velocity ranges: 
One is the range $3\lesssim V_{\rm LSR}\lesssim6$ \kms 
and the other is the range $9\lesssim V_{\rm LSR}\lesssim12$ \kms. 
The \co emission is weak in the range
$6\lesssim V_{\rm LSR}\lesssim9$ \kms where the \hco emission is prominent.
This indicates that there is a velocity component at \Vlsr$\,\simeq 7$ \kms,
and the optically thick \co emission line is heavily self-absorbed
while the optically thinner \hco emission appears as emission,
which is obvious in the position-velocity (PV) diagrams shown in
Figures \ref{fig:V0} (a) and (b) 
together with the intensity-weighted mean velocity map of the \hco emission in Figure \ref{fig:V0} (c). 
Note that, in addition to the velocity component at $\sim7$ \kms,
there are also distinct components at $\sim5$ and $\sim10$ \kms detected in \hco
just around the \Hii region.
In the next subsection, we discuss the details of the velocity structures
of the observed region.


\subsection{Four Velocity Components} \label{sec:kinematics}

A careful inspection of the \co and \hco data infers that the observed region consists of at least 
four velocity components at $V_{\rm LSR}\simeq3$, 5, 7, and 10 \kms.
As summarized in Table \ref{tab:components}, the $\sim7$ \kms component represents the systemic
velocity of the entire W40 region, 
and the $\sim5$ and $\sim10$ \kms components are likely to originate from dense gas 
located along the expanding shell of the \Hii region. 
The dense gas corresponds to Clump 1 ($\sim5$ \kms) and Clumps 5 and 6 ($\sim10$ \kms)
detected in \hco that we will describe in Sections \ref{sec:dense} and \ref{sec:model}.
The $\sim3$ \kms component is weak and shows up in a very limited region, and it might be unrelated to W40.
As explained  in the following, the four velocity components can be recognized in the \co and \hco spectra.

\begin{figure*}
\begin{center}
\includegraphics[scale=0.85]{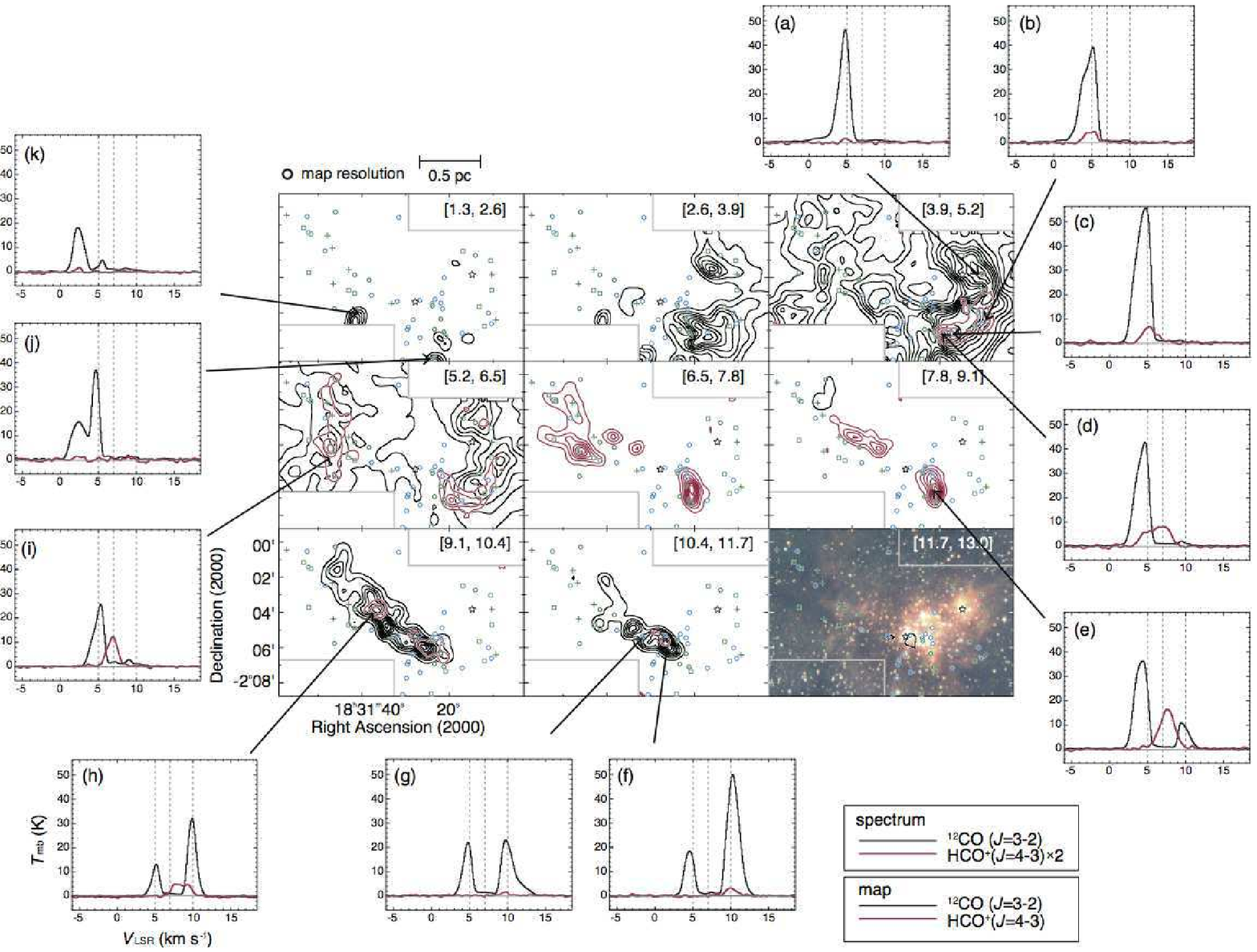}
\caption{
Channel maps of the \co (black contours) and \hco (red contours) emission lines in the velocity range
$1.3 \leqq V_{\rm LSR} \leqq 13.0$ with the 1.3 \kms step.
The velocity range in units of \kms used for the integration is indicated in the top right corner of each panel. 
The lowest contours and the contour intervals are 5 K \kms for the \co maps, and 1 K \kms for the \hco maps.
The 2MASS three-band composite image ($JHK$s) is overlaid in the last panel.
The symbols for the YSOs are the same as in Figure \ref{fig:iimap}.
Representative \co (black lines) and \hco (red lines) spectra taken at some positions
are also shown in panels (a) to (k). 
The \hco spectra are scaled up by a factor 2.
The vertical broken lines indicate the typical velocities of the three
components at $V_{\rm LSR} \simeq $5 \kms, 7 \kms, and 10 \kms in Table \ref{tab:components}.
\label{fig:spe}}
\end{center}
\end{figure*}

In Figure \ref{fig:spe}, we show examples of the \co and \hco spectra observed at  some positions
mainly around the YSOs together with the velocity channel maps with a coarser velocity interval (1.3 \kms)
than Figure 2.
The four components can be seen in some of the spectra. 
For example, the spectra in panels (a)--(c) sampled in the west of the \Hii region 
consist only of the $\sim5$ \kms component, 
and those in panels (f) and (g) sampled in the south of the \Hii region exhibit the $\sim10$ \kms component as well.
Note that the $\sim5$ and $\sim10$ \kms components
are detected both in \co and \hco only around the \Hii region.
In addition, the \hco spectra in panels (e) and (i) consist of the $\sim7$ \kms component. 
The $\sim3$ \kms component is detected both in \co and \hco 
and is seen only around the positions shown in panels (j) and (k).

Compared with the 2MASS image in Figure \ref{fig:spe}, the $\sim5$ and $\sim10$ \kms components 
exhibit intense \co emission, and they are likely to delineate the expanding shell around the \Hii region.
The $\sim7$ \kms component is well detected in \hco and is likely to represent the systemic velocity of the entire W40 region. 
As mentioned earlier, the absence of the intense \co emission for the $\sim7$ \kms component indicates that 
there is a large amount of cold gas in the foreground of the \Hii region, 
which should cause the heavy absorption in \co around this velocity. 
Actually, the \co emission line over the velocity range 
$6 - 9$ \kms have a brightness temperature of only $T_ {\rm mb}\simeq 2$ K while the optically
thinner \hco emission line is brighter (up to $\sim 8$ K) as seen in Figure \ref{fig:spe} (e). 

The above results are consistent with those obtained through earlier observations 
\citep[e.g.,][]{Crutcher, Pirogov}.
Molecular observations done by \cite{Pirogov} covers the arc-shaped structure just around the \Hii region,
and they detected the two velocity components at $\sim5$ and $\sim7.5$ \kms. 
\cite{Crutcher} made \co$(J=1-0)$ observations with a lower angular resolution (HPBW$ =2\farcm6$),
and they detected the \co emission at $\sim5$ and $\sim10$ \kms. 
The author also found an absorption line at $\sim7$ \kms in the OH spectra
obtained by additional observations with the NRAO 43 m telescope (HPBW$=18\arcmin$).
Based on these data, \cite{Crutcher} concluded that 
W40 has a single component with a peak velocity at $V_{\rm LSR}\simeq 7$ \kms, 
which is extended in the wide velocity range $3-12$ \kms 
and is self-absorbed at around 7 \kms by colder gas in the foreground. 
We agree with their interpretation 
that W40 is surrounded by the $\sim7$ \kms component. 
In addition, we found the existence of 
two other velocity components at $\sim5$ and $\sim10$ \kms around the \Hii region,
probably originating from the dense gas located along the expanding shell,
which are detected both in \co\3 and \hco\4 as seen in the PV diagram in Figure \ref{fig:V0} 
as well as in the spectra in Figures \ref{fig:spe} (b) and (f).


\subsection{Search for Outflows} \label{sec:Outflow}

Looking at the \co spectra in Figure \ref{fig:spe}, 
there are some cases in which high velocity wings are obviously seen
in spite of the huge complexity in the velocity field.
For example, the \co spectrum in Figure \ref{fig:spe} (a) 
exhibits a blue-shifted wing at  \Vlsr $\,\leq3$ \kms, 
and the one in Figure \ref{fig:spe} (g) exhibits a red-shifted wing at \Vlsr $\,\geq12$ \kms. 
These wings are very likely to be tracing real high velocity gas,
not mere Gaussian tails of the intense \co spectra of the $\sim5$ and $\sim10$ \kms
components. As detailed in the Appendix \ref{sec:gaussian_fit}, most of the \co spectra
in the observed region can be well fitted by simple Gaussian functions
except for the apparent high velocity wings as seen in Figure \ref{fig:spe} (a) and (g).

Because the high velocity wings are often seen in the \co spectra along the dense condensations
traced in \hco both around the \Hii region and the filamentary structure, we suggest that
they are either the high velocity gas blown away by the expansion of the \Hii region or
the outflowing gas driven by YSOs forming in the dense condensations.
At the moment, there is no way to distinguish these two effects reliably only with
the present dataset, and both of them should be actually at play.
Modeling the expected signature of instabilities in high velocity gas from \Hii region expansion 
is beyond the scope of the current study. 
For the purpose of this study, we interpret apparent high velocity wings 
on the principal velocity components of \co as evidence of potential ``candidate" outflows 
from YSOs and derive their parameters accordingly. 
Though a certain fraction of the high velocity wings can be due to the expansion of the \Hii region,
there are some high velocity wings found in the vicinity of YSOs (e.g., ``B2" and ``R2" in the following)
including the one located away from the \Hii region (e.g., ``R4") which are likely to be promising
outflows from YSOs.

\begin{figure*}
\begin{center}
\includegraphics[scale=1.]{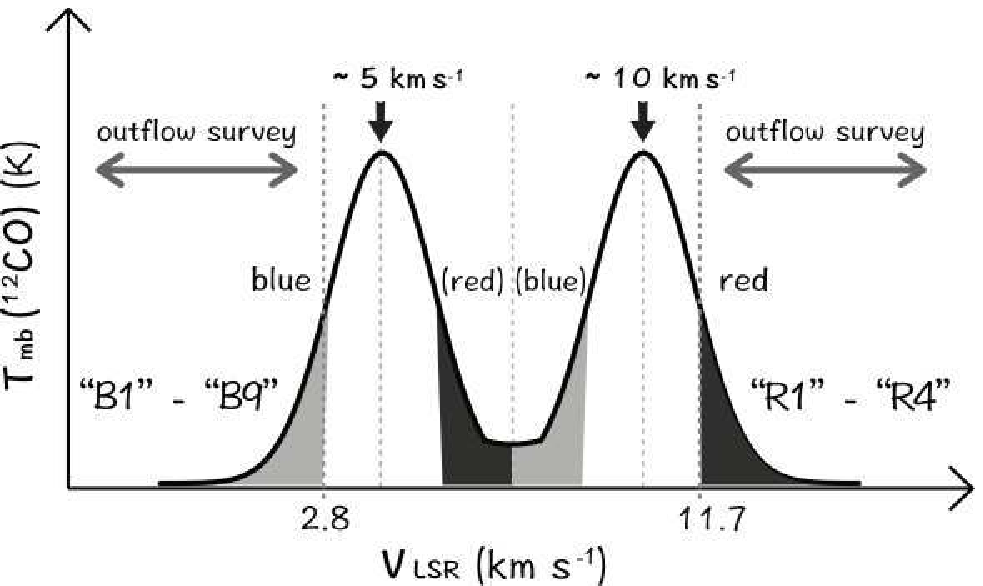}
\caption{
Schematic illustration to search for the high velocity wings of outflows in the \co spectra. 
In the W40 region, 
there are three main velocity components at $V_{\rm LSR}\simeq5$, 7, 10 \kms (Table \ref{tab:components})
, but the heavy absorption by the $\sim7$ \kms component
precludes the detection of the high velocity wings 
over the velocity range $6\lesssim V_{\rm LSR} \lesssim 9$ \kms. 
We therefore searched for the \co emission in the velocity ranges
$-3.2\leqq V_{\rm LSR}\leqq2.8$ \kms
and
$11.7\leqq V_{\rm LSR}\leqq14.5$ \kms
to find candidates of the blue- and red-shifted wings of the outflows, respectively.
The \co intensity integrated over these velocity ranges are shown 
by the blue and red solid contours in Figure \ref{fig:outflow}. 
Outflow candidates originating from the $\sim5$ \kms and $\sim10$ \kms components
may show symmetric wing components, and so, for a reference, 
we further investigated possible wing components contained in the velocity ranges
$5.7\leqq V_{\rm LSR}\leqq8.4$ \kms
and 
$8.5\leqq V_{\rm LSR}\leqq9.2$ \kms.
The \co intensities integrated over these velocity ranges are shown by the red and blue broken contours
in Figure \ref{fig:outflow}.
\label{fig:model1}}
\end{center}
\end{figure*}

Since the dense gas consists of the $\sim5$ \kms, $\sim7$ \kms, and $\sim10$ \kms components, 
we searched for outflow candidates blowing out from these components. 
For the $\sim$3 \kms component, we did not find any high velocity wings in the \co spectra, and thus
we disregarded this component.
In order to carry out a systematic search for outflows, 
we set constant velocity ranges to find high velocity wings in the \co spectra. 
As illustrated in Figure \ref{fig:model1}, we decided to use the velocity ranges
$-3.2\leqq V_{\rm LSR}\leqq2.8$ \kms and $11.7\leqq V_{\rm LSR}\leqq14.5$ \kms 
for the blue- and red-shifted high velocity wings of outflows, respectively. 
Basically, we can carry out a reliable search for outflow candidates in \co only in these velocity ranges,
because the heavy absorption by the $\sim7$ \kms component precludes the detection of 
the faint wing emission over the range $6 \lesssim V_{\rm LSR} \lesssim9$ \kms.  
However, for a reference, 
we will also investigate wing emission possibly associated with the $\sim5$ and $\sim10$ \kms
components in the range $5.7\leqq V_{\rm LSR}\leqq8.4$ \kms and $8.5\leqq V_{\rm LSR}\leqq9.2$ \kms,
because, in the vicinity of the \Hii region, \cite{Zhu} found a bipolar
outflow associated with the $\sim5$ \kms component, though we now realize
that its red-shifted wing emission should be contaminated by the heavy absorption.

\begin{figure*}
\begin{center}
\includegraphics[scale=.75]{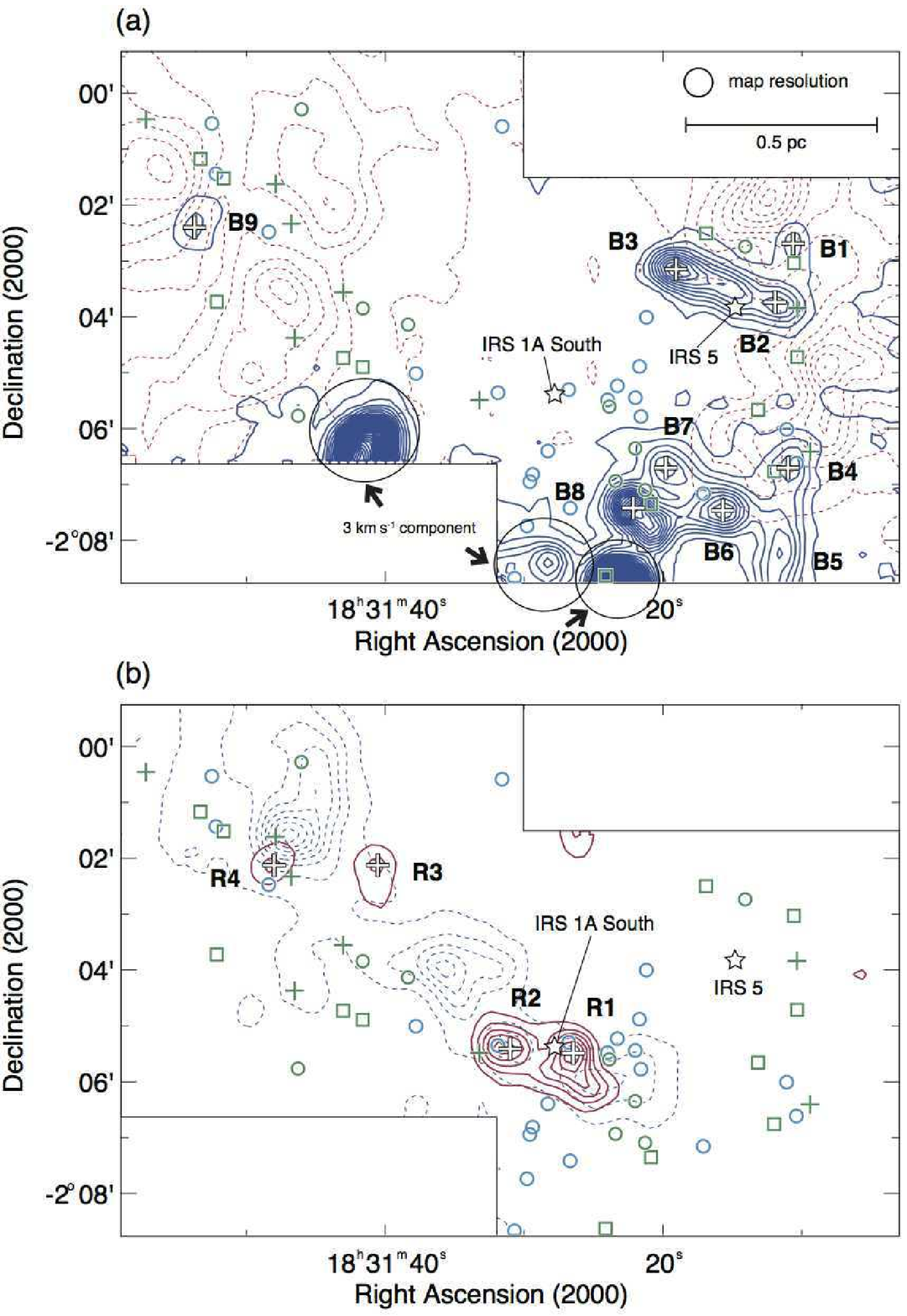}
\caption{
(a) Locations of blue lobes of the outflow candidates which are labeled ``B1"--``B9".
White plus signs denote the peak positions of the lobes.
Blue solid and red broken contours represent the \co intensity
integrated over the velocity range $-3.2\leqq V_{\rm LSR}\leqq2.8$ \kms 
and $5.7 \leqq V_{\rm LSR} \leqq 8.4$ \kms, respectively. 
The former contours are to search for the blue lobes of outflows, and the latter contours
are made for a reference (see text and Figure \ref{fig:model1}). 
The lowest contour and the contour interval are 1.2 K \kms,
and symbols for the YSOs are the same as in Figure \ref{fig:iimap}.
Clumps indicated by thin solid circles with arrows 
are due to the distinct velocity component (at $\sim3$ \kms) unrelated to the
lobes of the outflow candidates.
(b) Same as (a), but for red lobes of the outflow candidates.
Identified lobes are labeled ``R1"--``R4".
Red solid and blue broken contours represent the \co intensity
integrated over the velocity range $11.7 \leqq V_{\rm LSR} \leqq 14.5$ \kms 
and $8.5 \leqq V_{\rm LSR} \leqq 9.2$ \kms, respectively. 
The former contours are to search for the red lobes of outflows, and the latter contours
are made for a reference (see text and Figure \ref{fig:model1}). 
\label{fig:outflow}}
\end{center}
\end{figure*}

Figure \ref{fig:outflow} shows the \co intensity distributions integrated over the
velocity ranges mentioned in the above. 
The overall distributions of the high velocity gas in the figures show a great complexity
similar to what has been reported for the Circinus cloud where the complex outflow lobes are found to
be a mixture of some smaller outflows \citep[e.g.,][]{Dobashi1998,Bally1999,Shimoikura2011},
which is likely the case also for W40.
In addition, the \co distributions in the figures are contaminated by some distinct velocity 
components (at $V_{\rm LSR}\simeq 3$ \kms) unrelated to the outflows. 

The \co distributions on the sky are too complex to separate them reliably into single
outflow candidates only from the present dataset.
In this paper, we thus decided to pick up the local peaks in the \co intensity map in Figure \ref{fig:outflow}
if they are not due to the distinct velocity components but are very likely due to outflows showing
a wing-like feature in the \co spectra. 
As a result, we found 13 local peaks as indicated in Figure \ref{fig:outflow}.
For simplicity, we will regard each of the 13 local peaks as
the peak position of a single high velocity lobe and we will consider them to be candidates of outflows in W40.
There are nine blue lobes and four red lobes. 
As summarized in Table \ref{tab:outflow1}, we call the blue lobes ``B1"--``B9", and the red lobes ``R1"--``R4". 

To measure some physical properties, we defined the surface area of the outflow lobes $S_{\rm lobe}$
at the 1.2 K \kms (corresponding to the $\simeq9 \sigma$ noise level) contour level. 
In the case that two or more lobes are attached to each other at this threshold value, 
we divided them into single lobes at the valley in the contour maps of the \co intensity (Figure \ref{fig:outflow}). 
We also attempted to find driving sources of the outflow lobes. 
It is however also difficult to identify the driving sources reliably,
and thus we simply picked up YSOs located within $S_{\rm lobe}$ defined in the above
as possible candidates for the driving sources.
There are nine outflow candidates (B1, B2, B4, B6, B7, B8, R1, R2, and R4) 
with at least one candidate of the driving source, 
and there are four outflow candidates (B3, B5, B9, and R3) which could not be assigned to 
any known sources.

We summarize the main features of the outflow lobes as follows.

{\bf  Blue lobes:}
Blue lobes are seen mainly around the \Hii region.
A candidate for the driving source of B1 is 1.2 mm source, 
which is found and classified as ``starless" by \cite{Maury}.
A jet-like blue lobes B2 and B3 elongated in the East-West direction
are clearly seen in Figure \ref{fig:outflow}. 
B3 is the largest lobe located close to IRS 5.  
Multiple blue lobes B4--B8 are located to the south of the \Hii region.
B9 is a small lobe found in the filament structure.
B1--B8 correspond to an outflow discovered by \citet[][see their Figure 3]{Zhu}
with lower angular resolution ($80\arcsec$), which is resolved into smaller multiple
possible outflows in our map.

{\bf  Red lobes:}
Two red lobes R1 and R2 are detected in the vicinity of IRS 1A South, suggesting
that these outflow candidates possibly originate in the cluster.
Candidates of driving sources for these lobes are classified as Class I protostars \citep{Mallick}.
The lobes R3 and R4 are found in the filament structure.

\begin{figure*}
\begin{center}
\includegraphics[scale=.75]{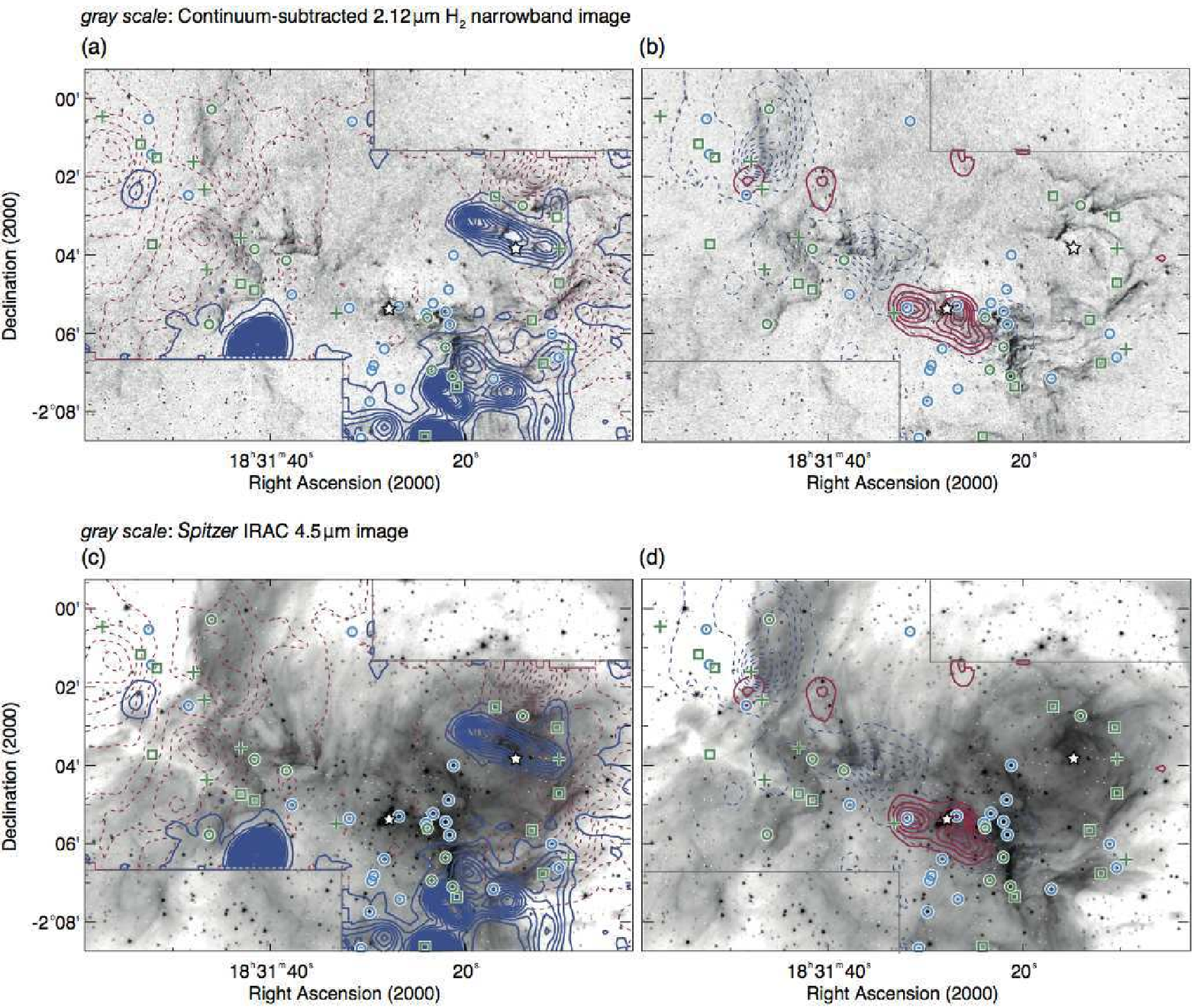}
\caption{
Comparison of the CO outflow candidates (contour) and other emission lines (gray scale).
(a) Locations of blue lobes of the outflow candidates (same as in Figure \ref{fig:outflow}(a))
overlaid with the continuum-subtracted 2.12 $\micron$ H$_{2}$ narrowband image obtained by \cite{Mallick}.
(b) Same as (a), but for red lobes (same as in Figure \ref{fig:outflow}(b)). 
(c) Locations of blue lobes of the outflow candidates overlaid with the {\it Spitzer} IRAC 4.5 $\micron$ image.
(d) Same as (c), but for red lobes. 
The symbols for the YSOs are the same as in Figure \ref{fig:iimap}.
\label{fig:outflow2}}
\end{center}
\end{figure*}

Finally, in Figure \ref{fig:outflow2}(a) and (b), we compared the distributions of the outflow candidates 
with the continuum-subtracted 2.12 $\micron$ H$_2$ narrowband image 
obtained by \cite{Mallick} to search for the possible counterparts of the CO outflow lobes. 
We also compared our map with the {\it Spitzer} IRAC 4.5 $\micron$ image
in Figure \ref{fig:outflow2}(c) and (d) to see whether vibrationally excited CO emission is detected
around the outflow candidates. 
However, we could not identify clear counterparts corresponding to the outflow lobes in these images, 
though there are some near/mid-infrared nebulosity in the vicinity.

According to the {\it Herschel} map (the left panel of Figure \ref{fig:iimap} (b)), 
the \Hii region is apparently expanding into the surrounding medium,
and some of the identified lobes (i.e., B2, B3, R1, and R2) are located 
in less dense regions at the edge of the \Hii region.
Furthermore, these lobes are associated
with the $\sim 5$ and $\sim10$ \kms components
which are likely to represent the expanding velocity of the shell
(see Section \ref{sec:model}).
For the lobes found just around the \Hii region, we cannot
distinguish if the observed high velocity gas is due to real outflows
from YSOs or is merely tracing the expanding shell.
In addition to the above possibility, we should also note that some of
the identified lobes might not originate from a single YSO (see Figure \ref{fig:outflow}),
but originate from the cluster in the \Hii region.
All of these ambiguities arise from the complex velocity filed in W40,
especially the heavy absorption by the $\sim7$ \kms component.
In this paper, we will regard the 13 blue- and red-shifted high velocity lobes 
to be candidates of outflows in this region and will estimate some outflow parameters
in Section \ref{sec:parameters}. 


\subsection{Properties of Outflow Candidates \label{sec:parameters}}

In order to estimate the masses of the outflow candidates, 
we assumed that the \co molecules are in the local thermal equilibrium (LTE)
and also that the \co emission is optically thin at all wing velocities
to calculate the column densities in the outflow lobes for
a uniform excitation temperature of $T_{\rm ex}=30$ K.
We then estimated the mass, momentum, and energy of the outflow lobes.
Details of the calculations of the physical parameters are described in Appendix \ref{sec:Outflow2}, 
and the results are summarized in Table \ref{tab:outflow1}. 

\cite{Zhu} suggested a total mass of 0.48 $M_{\sun}$ for the blue lobe of the outflow they found
which corresponds to our B1 to B8,
on the assumption of a distance of 600 pc and a mean molecular weight of 2.72 
(cf., we assumed 500 pc and 2.4, respectively).
The sum of the corresponding blue lobes in our Table \ref{tab:outflow1} amounts
only to $\sim0.11$ $M_\sun$ which would be rescaled to $\sim0.18$ $M_\sun$ for the same 
assumptions as theirs. The difference between their and our estimates is mainly due to
the contamination of the distinct velocity component at $V_{\rm LSR}\simeq 3$ \kms (see Section \ref{sec:kinematics})
which was not resolved in their observations, and also due to the fact that we defined the 
surface area of the outflow lobes at a rather high threshold value of 1.2 K \kms
(see Section \ref{sec:Outflow}).

As seen in the 8th column of Table \ref{tab:outflow1}, 
we found that the masses of the individual lobes are in the range from 
$\sim0.001$ $M_{\sun}$ to $\sim0.03$ $M_{\sun}$. 
The YSOs that can be the possible driving sources of these outflow candidates have bolometric luminosities
$L_{\rm bol}\simeq 4.2-6.6\,L_\sun$ \citep[see their Table 1]{Maury}. 
According to a statistical study of outflows by \cite{Wu}, 
outflows associated with low-mass YSOs with $L_{\rm bol} \simeq1-10 \,L_{\sun}$
have a mass range of $0.01$ $M_{\sun}$ to a few $M_{\sun}$. 
The masses of outflow candidates we found in W40 are much smaller than the average mass of outflows
associated with low-mass YSOs, and are close to the lower end in the plots
of the outflow masses versus the bolometric luminosity of the driving sources
reported by \cite{Wu} (see their Figure 5). 

Here we estimate the possible errors in the derived parameters caused by our assumptions.
First, we assumed the constant excitation temperature $T_{\rm ex}=30$ K.
An error arising from this assumption should be small, because the outflow masses are
insensitive to $T_{\rm ex}$, and they would fluctuate only by $\sim$20$ \%$ at most within the variation
of $T_{\rm ex}$ in the mapped region ($20<T_{\rm ex}<65$ K).
Second, we assumed that wing components in the \co\3 emission line are optically thin ($\tau\ll 1$),
but they might be moderately optically thick. For example, \citet{Dunham} investigated 17 outflows associated with
low mass stars with the \co\3 and $^{13}$CO \3 emission lines, and found the optical depths of the \co\3 emission line
varies in the range $\tau\simeq 1-10$ with an average value of $\sim2.8$. 
If we adopt their average value, the inverse of the escape probability in Equation (\ref{eq:g}) would be $\beta^{-1}=3.0$.
This infers that our optically thin assumption may cause an underestimation in the mass estimate by a factor of a few. 
Finally, it is important to remark that the parameter $V_{\rm wing}$ 
(i.e., the velocity range exhibiting the high velocity wings) 
in Equations (\ref{eq:Nco})--(\ref{eq:Energy})
should be highly underestimated because of the heavy absorption in the 
velocity range $6 \lesssim V_{\rm LSR} \lesssim9$ \kms, 
and hence our estimate for the mass, energy, and momentum of the outflows should be highly underestimated.

To summarize our outflows survey, we found that 13 candidates of outflow lobes showing
wing-like emission in the \co spectra. 
However, we should note that our estimates of the outflow parameters such as 
the mass, momentum, and energy should give only the lower limits to the actual values 
mainly because of the heavy absorption at $\sim7$ \kms and the optically thin assumption. 
For most of the outflow candidates identified, we cannot even pair red and blue lobes,
because of their complex distributions \citep[e.g.,][]{Arce,Offner,Bradshaw}. 

\color{black}

\subsection{Clumps Detected in HCO$^{+}$} \label{sec:dense}

\begin{figure*}
\begin{center}
\includegraphics[scale=.8]{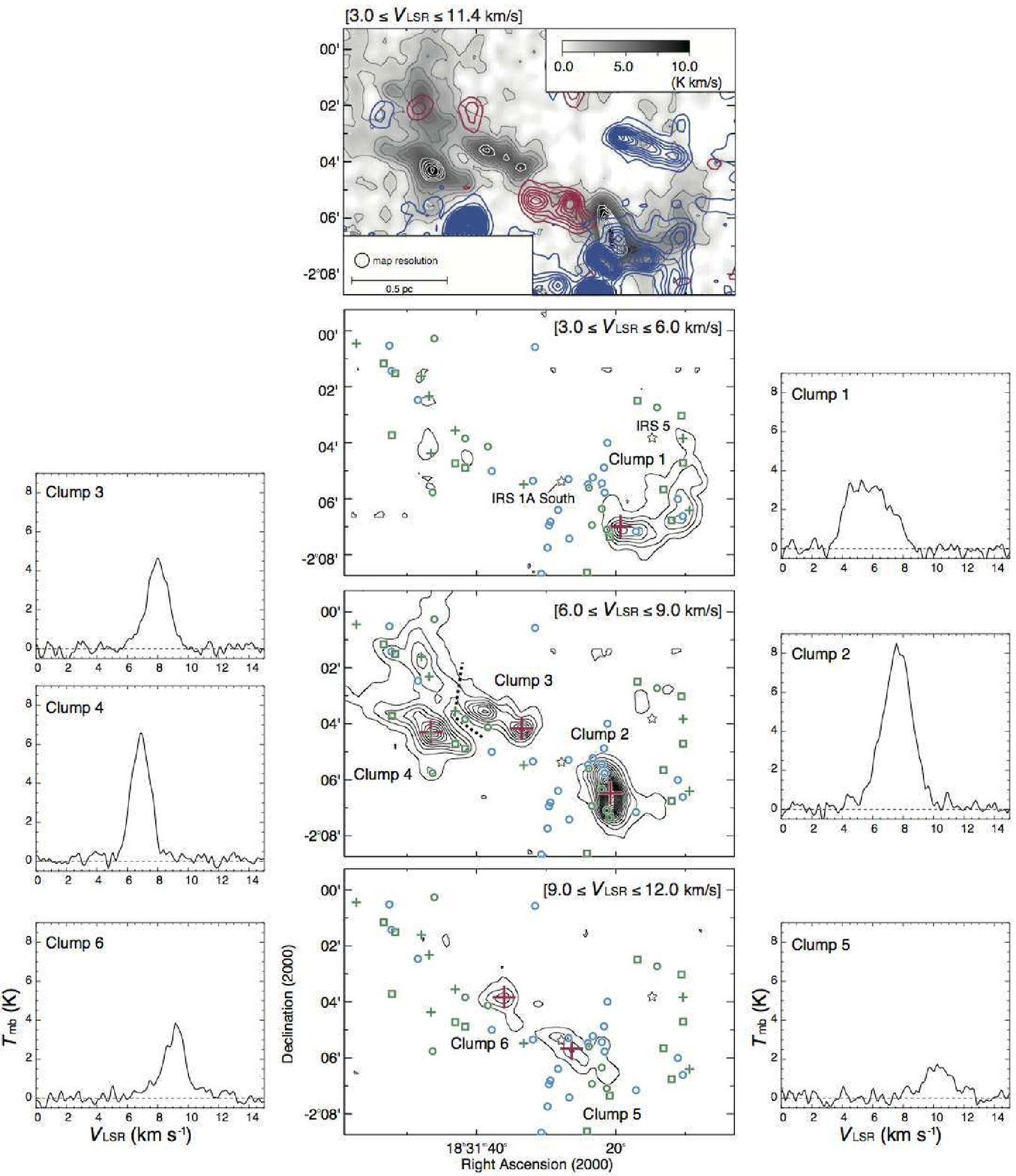}
\caption{
Top panel shows the total integrated intensity map of HCO$^{+}$ same as Figure \ref{fig:iimap}(b) 
overlaid with the candidate outflow lobes. 
The other three panels below the top panel display the channel maps for the
$\sim5$ \kms,
$\sim7$ \kms, and
$\sim10$ \kms components
integrated over the range
$3 \leqq V_{\rm LSR} \leqq 6$ \kms,
$6 \leqq V_{\rm LSR} \leqq 9$ \kms, and
$9 \leqq V_{\rm LSR} \leqq 12$ \kms, respectively.
Six identified clumps are indicated in these panels. 
The lowest contours and the contour intervals in the three panels are 1 K km s$^{-1}$. 
Peak positions of the six clumps are indicated by red plus signs. 
The symbols for the YSOs are the same as in Figure \ref{fig:iimap}. 
The \hco spectra sampled at the peak positions are also shown.
\label{fig:HCOspe}}
\end{center}
\end{figure*}

The \hco \4 emission line is a good tracer of cold dense gas for its high critical density $>10^{6}$ cm$^{-3}$\citep[e.g.,][]{Girart}. 
As stated in Section \ref{sec:kinematics}, the radial velocities of the \hco emission
can be classified to the three velocity components at $\sim5$, 7, and 10 \kms
in Table \ref{tab:components}, and they can be mostly well separated at $V_{\rm LSR}=$6.0 and 9.0 \kms.
We show the \hco integrated intensity maps divided at these velocities in Figure \ref{fig:HCOspe}.
In order to investigate physical properties of the dense gas detected in \hco, we will define clumps
based on these maps. We will regard a condensation enclosed by the lowest contours 
(1 K \kms$ \simeq 6\, \sigma$) having
the peak integrated intensity greater than 2 K \kms in each map as one clump. 
There are five condensations satisfying the criteria. 
However, a large condensation in the $6\leqq V_{\rm LSR}\leqq 9$ \kms range 
exhibits two slightly different velocity components by $\sim 1$ \kms, and thus we divided them
into two at the boundary of the two velocity components as labelled Clumps 3 and 4
in Figure \ref{fig:HCOspe}.
As a result, we identified six clumps in the \hco data. We show the \hco spectra sampled at their peak positions 
in Figure \ref{fig:HCOspe}, and also show the line parameters (i.e., the peak brightness temperature, line width, and peak velocity)
obtained by fitting the spectra with a single Gaussian function in Table \ref{tab:clump}.

Based on the \hco data, we will derive molecular masses of the clumps
on the assumption of the LTE using Equations (\ref{eq:NX}) and (\ref{eq:g}) 
for $X=$HCO$^+$ (see Appendix \ref{sec:Outflow2}). 
According to JPL Catalog\footnote{http://spec.jpl.nasa.gov/} and Splatalogue\footnote{http://www.splatalogue.net}, 
the constants $B_{0}$, $\mu$, and $E_{u}$ 
are 44.5944 GHz, 3.888 debye, and 29.7490 cm$^{-1}$, respectively, for the \hco\4 emission line.
We list these constants in Table \ref{tab:const}.

In order to derive the mass of the clumps, we estimated $T_{\rm ex}$ for \hco
comparing our \hco\4 data with the \hco$(J=1-0)$ data measured
by \cite{Pirogov} toward some limited positions (see their Table 4). 
For these positions, we smoothed our \hco\4 spectra to the same angular resolution
as the \hco$(J=1-0)$ spectra ($\sim45\arcsec$), and applied 
the non-LTE radiative transfer code RADEX \citep{Van2007}.
Results of the analyses infer $T_{\rm ex} \gtrsim 30$ K
to reproduce the hydrogen number density $n({\rm H}_2)\simeq 1 \times 10^6$ cm$^{-3}$
found by \cite{Pirogov} for these positions. 
This value of $T_{\rm ex}$ is consistent with the temperatures
of dense gas around the \Hii region measured by \cite{Pirogov} using
NH$_3$ and dust continuum data
($16 - 40$ K, see their Table 7). In this paper, we therefore adopt $T_{\rm ex}=30$ K
throughout the observed region.

If we assume that the emission line is optically thin ($\tau \ll 1$), 
the \hco column density can be derived by the following equation for $T_{\rm ex}=30$ K,
\begin{equation}
\label{eq:Nhco}
N_{\rm HCO^+}=4.53\times10^{11}\int_{V_{\rm range}} T_{\rm mb}^{\rm HCO^+}(v)~{\rm d}v ~~~~{\rm cm^{-2}}
\end{equation} 
where the term $\int T_{\rm mb}^{\rm HCO^+}(v)~{\rm d}v $ is in units of K \kms.
The velocity range for the integration $V_{\rm range}$ is the same as shown in Figure \ref{fig:HCOspe}, i.e., 
$3\leqq V_{\rm LSR}\leqq 6$ \kms, $6\leqq V_{\rm LSR}\leqq 9$ \kms, and $9\leqq V_{\rm LSR}\leqq 12$ \kms
for clumps at $\sim 5$ \kms (Clump 1), $\sim 7$ \kms (Clumps 2--4), and 
$\sim 10$ \kms (Clumps 5 and 6), respectively.

Assuming the fractional \hco abundance ratio to be $N_{\rm HCO^+}/N_{\rm H_2}=2.5 \times 10^{-10}$
found by \cite{Pirogov} through comparison of the \hco ($J=1-0$) emission line and 1.2 mm continuum data toward our Clump 1, 
the total mass of the clumps can be estimated by the following equation,
\begin{equation}
\label{eq:ClumpMass}
M_{\rm clump}=2.96\times10^{-6}\,D^2 {\int_{S_{\rm clump}} \int_{V_{\rm range}}T_{\rm mb}^{\rm HCO^+}}(v){\rm d}v{\rm d}S~~~ M_\sun
\end{equation} 
where the term ${\int \int T_{\rm mb}^{\rm HCO^+}} {\rm d}v {\rm d}S$ is in units of K \kms arcmin$^2$ 
and $D$ is the distance to the clumps in units of pc ($D=500$ pc for W40).
We defined $S_{\rm clump}$ is the surface area of the clump at the 1 K \kms contour level.

Using Equations (\ref{eq:Nhco}) and (\ref{eq:ClumpMass}), we estimated the total mass of the clumps.
The results are presented in Table \ref{tab:clump}. As seen in the table, the identified clumps have a masses
ranging from $\sim2$ to 26 $M_\sun$ with a mean value of $\sim12$ $M_\sun$.

Here, we should note that our mass estimates depend on the adopted $T_{\rm ex}$.
The derived masses should increase for lower $T_{\rm ex}$, and they would change by $\sim50$ $\%$
at most if $T_{\rm ex}$ varies in the range $16 - 40$ K as measured by \cite{Pirogov}.

\cite{Pirogov} suggested 
a total mass of 8.1 $M_\sun$ (the sum of the masses of their clumps 2 and 3)
for a part of our Clump 2 from the 1.2 mm continuum data. 
We estimate the mass contained in the same area measured by Pirogov et al. to be 9 $M_\sun$.
The mass derived from our \hco data is therefore in close agreement with that derived by \cite{Pirogov}.
However, on a larger scale, \cite{Maury} suggested a much larger mass than our estimate.
They estimated the total mass of a larger region equivalent to our entire observed area
to be 310 $M_\sun$ through the {\it Herschel} column density map.
They assumed a distance of 260 pc, and thus the mass should be 
rescaled to $\sim1150\,M_\sun$ for our assumed distance (500 pc). 
We obtain only $\sim110\,M_\sun$ as the total mass of the observed region, which is only $10\%$
of their mass. This is probably due to the fact that the dust emission (detected by {\it Herschel})\
traces all of the dust along the line-of-sight, while \hco traces only very high density regions
because of its high critical density ($>10^{6}$ cm$^{-3}$). 
Moreover, as suggested by \citet{Girart}, \hco\4 is not likely to be thermalized, and therefore 
the masses derived here are lower limits.

Finally, we attempt to find the possible association between the identified clumps
and the outflow lobes found in the previous subsections, though it is difficult to 
establish their definite association because of the enormous complexity in the velocity 
field for which we cannot identify even their driving sources reliably.
Here, we regard that the clumps and the outflow lobes are associated 
if their surface areas overlap each other. 
All of the outflow lobes except for B1 and B3 have an overlap with one or more clumps.
We will regard that B1 and B3 are associated with Clump 1, because weak \hco emission
is extending from this clump toward these outflow lobes. 
We also checked the systemic velocity $V_{\rm sys}$ for each of the lobes 
(see Appendix \ref{sec:Outflow2} and the 6th column of Table \ref{tab:outflow1}).
For B7 and B8, these $V_{\rm sys}$ are close to the radial velocity of Clump 2
and we will regard these lobes are associated with Clump 2. 
The results summarized in the last column in Table \ref{tab:clump}.

Comparing the identified clumps with the distributions of the YSOs in Figure \ref{fig:HCOspe}, 
It is interesting to note that some of the \hco clumps (e.g., Clump 1 and Clump 4) 
are associated with Class 0/I sources at their peak positions.
Moreover,
the clumps associated with the outflow lobes 
tend to have broader line widths (see Table \ref{tab:clump}), suggesting that 
the outflow lobes are influencing on the kinematics of the parent clumps. 
\cite{Nakamura2007} investigated theoretically the complex interplay 
between outflows induced by star formation and turbulence of the parent clumps,
and showed that outflows should play a significant role in altering the star-forming environments. 
The energy and momentum of the outflows found in this study appear too
small to give significant influence on the parent clumps, but again, 
we should note that our estimates for the outflow parameters may
suffer from a large uncertainty. 

\subsection{Three-Dimensional Model of W40} \label{sec:model}

\begin{figure*}
\begin{center}
\includegraphics[scale=.8]{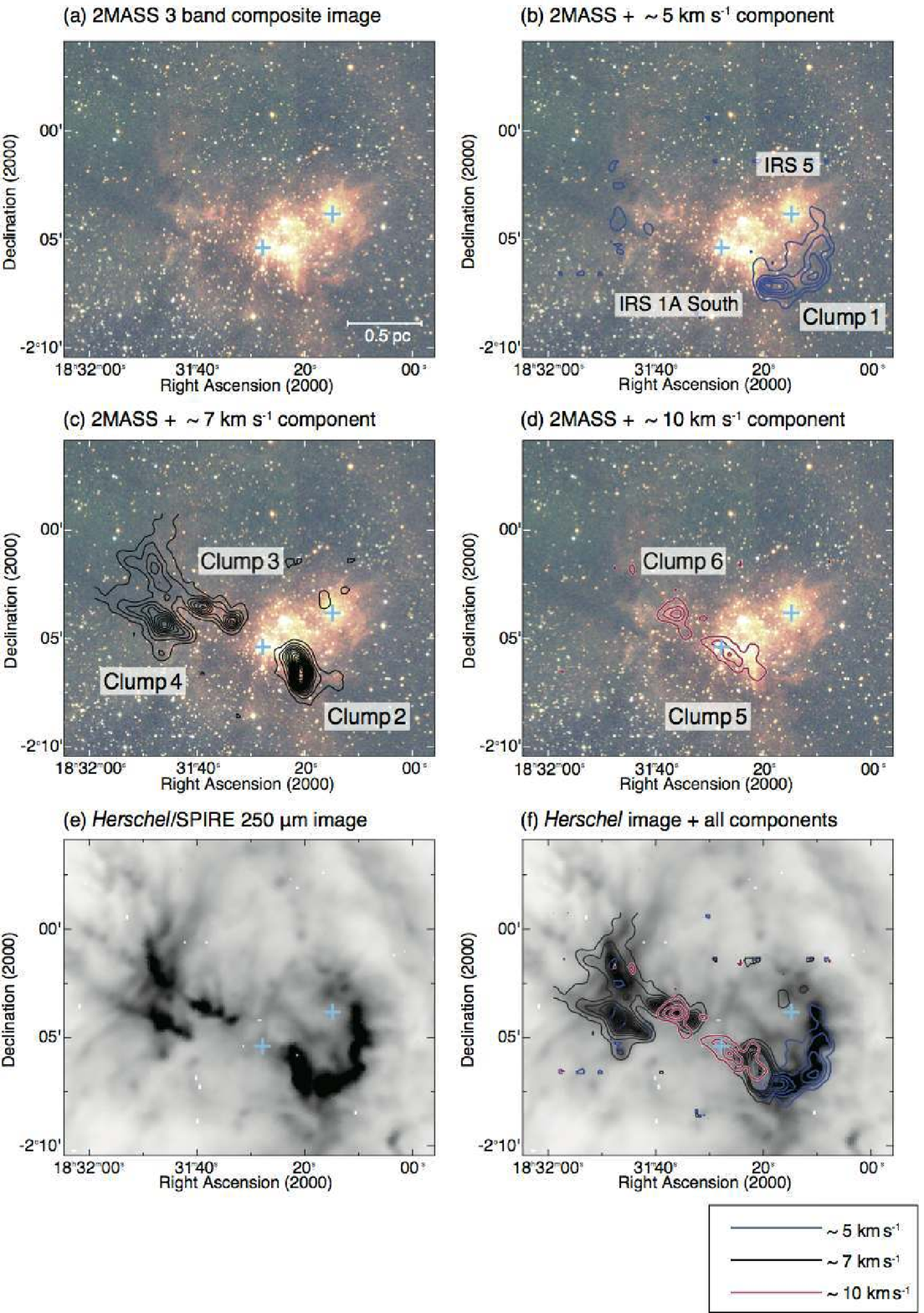}
\caption{
Distributions of dense gas traced in \hco and dust revealed by 2MASS and {\it Herschel}.
Panel (a) displays the three-colors ($JHK$s) composite image of 2MASS. Panels (b)--(d) display
the same 2MASS image, but overlaid with the \hco intensity maps (contours) integrated over the velocity range
$3 \leqq V_{\rm LSR} \leqq 6$ \kms (blue contours),
$6 \leqq V_{\rm LSR} \leqq 9$ \kms (black contours), and
$9 \leqq V_{\rm LSR} \leqq 12$ \kms (red contours), respectively.
Panel (e) displays the {\it Herschel} $250 \,\micron$ image. Panel (d) displays the same Herschel image,
but overlaid with the \hco intensity maps.
Blue plus signs denote the locations of IRS 1A South and IRS 5.
\label{fig:3comp}}
\end{center}
\end{figure*}

Here, we discuss the locations of
the identified clumps relative to the \Hii region and propose an updated three-dimensional model of W40.
Previous studies suggested that the \Hii region lies on the edge of a large molecular cloud
in a support of the blister model \citep[e.g.,][]{Zeilik,Vallee}. 
More recently, {\it Herschel} revealed the total dust distribution in W40
\citep[e.g.,][]{Andre, Konyves}.
We show the 250 $\micron$
image by {\it Herschel} and the three-colors ($JHK$s) composite image of 2MASS
in Figure \ref{fig:3comp} overlaid with our \hco channel maps 
($\sim 5$ \kms, $\sim 7$ \kms, and $\sim 10$ \kms).
Morphological coincidence between the molecular gas and the dust seen
in the far-infrared emission ({\it Herschel}) and in the near-infrared absorption (2MASS) is obvious.
It is noteworthy that Clumps 3 and 4 show up in the 2MASS image in the absorption,
indicating that these clumps are located in the foreground of the \Hii region.

Among the four velocity components summarized in Table \ref{tab:components},
the $\sim7$ \kms component is the primary component of molecular gas around W40.
Our Clumps 2--4 exhibit this velocity in \hco.
As seen in the PV diagram in Figure \ref{fig:V0},
there is a heavy absorption feature in the \co emission over the velocity range $6 \lesssim V_{\rm LSR}\lesssim9$ \kms.
\cite{Nakamura2011} have shown that the same absorption feature 
in \co is clearly seen around the Serpens South molecular cloud ($\sim20\arcmin$ to the west of W40) 
at the same velocity range. 
\cite{Dzib} measured the distance to Serpens Main to be approximately 415 pc through VLBI observations,
which is much further than previously thought \citep[e.g., 260 pc,][]{Maury}. 
On the other hand, the distance of W40 has not been determined precisely varying in the range 340--600 pc 
\citep[e.g.,][]{Rodney,Kuhn, Shuping}, 
and the W40 and Serpens South regions might be physically connected as suggested by some authors \citep[e.g.,][]{Bontemps}.
A more likely explanation for the absorption is that a colder cloud,
perhaps a part of the envelope surrounding both of the W40 
and Serpens regions,
lies in the foreground.
We thus regard that Clumps 3 and 4 are a part of the larger cloud surrounding these regions,
and that they are located in the foreground of W40.

In the \hco data, the other velocity components at $\sim5$ and $\sim10$ \kms are seen mostly around the \Hii region.
These components are very bright in \co ($T_{\rm mb}\gtrsim 40$ K),
indicating that they are probably interacting with the \Hii region being located close
to the ionization front.
This picture is consistent with the blister model suggested by \cite{Vallee} in which 
the \Hii region is expanding away from the dense molecular core located near the cloud surface.
We will develop the picture suggested by \cite{Vallee}.
The ionized gas associated with the \Hii region is expanding not only outward but also into the molecular cloud
pushing away the surrounding dense gas.
In this interpretation, Clumps 1, 2, and 5 should be part of the expanding shell. It is noteworthy that these clumps
have a radial velocity of $\sim 5$ \kms, $\sim7.5$ \kms, and $\sim10$ \kms. This infers that
they are the dense gas interacting with the \Hii region being swept up by the expansion of the ionization front.

\begin{figure*}
\begin{center}
\includegraphics[scale=.9]{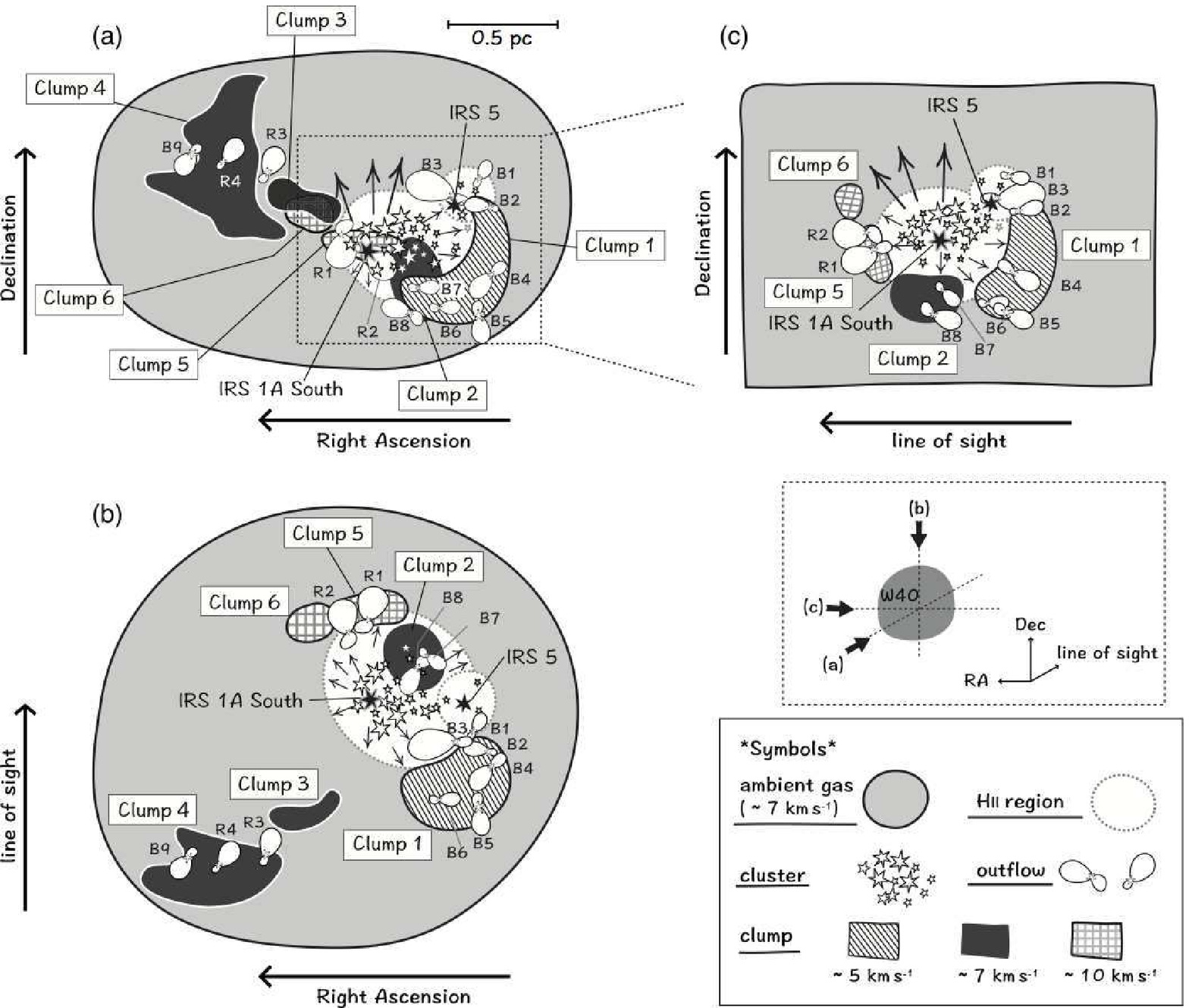}
\caption{
Schematic illustration of the proposed model for the three-dimensional geometry of the W40 region.
As indicated in the box drawn by the broken line,
panel (a) shows the view as is observed on the sky (e.g., see Figure \ref{fig:3comp}(b)--(d)),
and panel (b) shows the view observed from the declination axis orthogonal to the line-of-sight.
Panel (c) shows the close-up view around the \Hii region observed from the right ascension axis
(Clumps 3 and 4 are not drawn).
In the model, Clump 1 ($V_{\rm LSR}\simeq5$ \kms) is located on the near side of the expanding sphere of the \Hii region,
and Clumps 5 and 6 ($\sim10$ \kms) are located on the far side of the same sphere.
Clumps 3 and 4 ($\sim7$ \kms) are located in the foreground of the \Hii region.
\label{fig:model2}}
\end{center}
\end{figure*}

On the basis of the above results and discussion, we suggest a three-dimensional model of the clump distributions in W40,
which is summarized in the schematic illustration shown in Figure \ref{fig:model2}. 
In the following, we will discuss the locations of the individual clumps based on the \hco and
\co data and on their appearance in the 2MASS image (Figure \ref{fig:3comp}).
The molecular data provide us the information on the velocity and temperature of the clumps,
and the scattered light seen in the 2MASS image often infers the geometry between
the clumps and the \Hii region.

{\bf Clump 1:} In the 2MASS image, this clump is located very close to the \Hii region.
The rim of the clump facing to the \Hii region is bright, and the rest of the clump appears opaque (Figure \ref{fig:3comp} (b)).
The \co emission from this clump is very intense ($\sim53$ K, see Figure \ref{fig:spe} (c))
with a radial velocity of $\sim 5$ \kms which is blue-shifted by $\sim 2$ \kms compared to the systemic velocity 
of the entire W40 region ($\sim 7$ \kms). 
We thus suggest that the clump is located on the near side of the
expanding sphere of the \Hii region being heated and swept up toward us by the ionization front.

{\bf Clump 2:} This clump is the brightest in \hco among all of the clumps ( $\sim 8$ K, see Table \ref{tab:clump})
with a high density of $n({\rm H_2})\gtrsim10^6$ cm$^{-3}$ \citep{Pirogov}.
This clump overlaps with Clump 1 on the sky, but they can be clearly separated in velocity.
We cannot identify Clump 2 on the 2MASS image, but its counterpart can be recognized in the {\it Herschel}
250 $\micron$ image (see Figure \ref{fig:3comp}(f)), suggesting that Clump 2 is located behind Clump 1.
The radial velocity of Clump 2 is $\sim7.6$ \kms which is consistent with the systemic velocity, and thus the clump
might be located in the ambient background gas apart from the \Hii region. However, \cite{Pirogov} found a high
kinetic temperature of 40 K for this clump based on the non-LTE modeling for the CS $(J=2-1)$ and $(J=5-4)$
emission lines (see their Table 7), suggesting that Clump 2 is probably in the vicinity of the \Hii region.
The above observational features would be naturally accounted
for if we assume that the clump is interacting with the \Hii region like in the case of Clump 1, being swept
up toward the direction orthogonal to the line-of-sight, as illustrated in Figure \ref{fig:model2} (c).

{\bf Clumps 3 and 4:} These clumps are apart from the \Hii region on the sky, and
they can be easily identified in the 2MASS image as the opaque filaments (Figure \ref{fig:3comp} (c)).
The peak velocities of the \hco emission line of Clump 3 is $\sim 8$ \kms and that of Clump 4 is $\sim 7$ \kms
consistent with the systemic velocity of the W40 region.
These observational facts indicate that they are located in the foreground of the \Hii region as explained earlier.
However, we should note that Clumps 3 consists of two well defined peaks, and one of them in the south-west
appears as a small clump with the faint scattered light in the 2MASS image, not as the opaque region.
This part might be separated from the other peak with a different location along the line-of-sight,
though they are connected by the weak \hco emission.

{\bf Clump 5:} This clump has a radial velocity of $\sim10$ \kms which is red shifted by $\sim 3$ \kms
from the systemic velocity. The clump is very bright in \co ($\sim 50$ K, see Figure \ref{fig:spe} (f)),
suggesting that the clump is interacting with the \Hii region on the far side of the expanding sphere of the \Hii region. 
The extent of Clump 5 coincides with the bright region in the 2MASS image around IRS 1A South (Figure \ref{fig:3comp} (d)),
and we cannot recognize the shape traced in \hco in the image.

{\bf Clump 6:} This clump is located between the two local peaks of Clump 3 (at $\sim$8 \kms) on the sky,
but it has a clearly different velocity ($\sim9$ \kms).
There is a faint nebulosity around Clump 6 in the 2MASS image.
This clump is located rather far from the \Hii region on the sky, and it exhibits \co emission with a moderate
brightness temperature of $\sim30$ K (Figure \ref{fig:spe} (h)). We cannot decide the location
of this clump with a confidence, but we suggest that it may be located rather in the back of the \Hii region
because of its appearance in the 2MASS image as well as of its slightly red-shifted radial velocity
compared to the systemic velocity.

As illustrated in Figure \ref{fig:model2}, Clumps 1, 2, and 5 are likely to be located just around the \Hii region.
They are probably dense gas being compressed by the expansion of the \Hii region.
It is interesting to note that 10 outflow lobes found in the \co data are likely to be associated with these clumps.
This is suggestive of star formation triggered by the impact of expanding \Hii region 
driven by nearby massive young stars, indicating that the sequential star formation \citep[e.g.,][]{Elmegreen}
is taking place in this region. 

We also should note that three outflow lobes are likely to be associated with Clump 4
which is away from the \Hii region. It could be possible that the clump will be another site of cluster formation in W40. 
According to the study of 11 young clusters in the S247, S252, and BFS52 regions including the
Monkey Head nebulae \citep{Shimoikura2013}, four of the clusters actually consists of two or more smaller clusters
with a separation of $\sim1$ pc (e.g., see their Figure 11(c)). Such multiple cluster formation in a limited region
may not be rare, because there are some other young clusters showing multiplicity
\citep[e.g., the S235 region,][]{Kirsanova}. Though the relation between Clump 4 and
the \Hii region is unclear, the dust distributions on a larger scale indicate that there are two outstanding condensations of dust
around Clump 4 and the \Hii region (e.g., see Figure \ref{fig:iimap}(b)), suggesting that the W40 region might be a site of
sequential cluster formation.


\section{CONCLUSIONS}
We have carried out \co\3 and \hco\4 observations around the \Hii region W40 with the ASTE 10 m telescope
and identified CO outflows and dense clumps in this region. 
We have also observed the region with molecular lines at 45 GHz
such as SiO ($J=1-0$, $v=0$) and CCS ($J_{N}$=4$_{3}-3_{2}$) using the 45 m telescope at NRO. 
We summarize the main conclusions of this paper in the following points (1)--(4):

(1) Intense \co (up to $T_{\rm mb}\simeq50$ K) and \hco emission (up to $\sim 8$ K) lines were
detected around the W40 \Hii region. The velocity field in the region is highly complex, showing at least four 
distinct velocity components at $V_{\rm LSR} \simeq 3$, 5, 7, and 10 \kms.
The $\sim7$ \kms component represents the systemic velocity of the region
which causes heavy absorption in the \co spectra over the velocity range $6\lesssim V_{\rm LSR} \lesssim 9$ \kms.
The $\sim5$ and $\sim10$ \kms components exhibit high \co temperature ($\gtrsim40$ K),
and observed mainly around the \Hii region. 
The $\sim3$ \kms component is faint, and might be unrelated to W40.

(2) The SiO, CCS, HC$_{3}$N, and HC$_{5}$N emission lines were not detected at the present sensitivity.
We estimated the upper limits of the column density of CCS and its fractional abundance of W40 to be 
$\sim1.1 \times 10^{13}$ cm$^{-2}$ and $\sim 10^{-9.25}$, respectively.

(3) Based on the \co\3 data, we made a survey for molecular outflows
over the observed region. We searched for high velocity outflow lobes
in the velocity range $V_{\rm LSR}\leqq 2.8$ \kms for the blue lobes
and $V_{\rm LSR}\geqq 11.7$ \kms for the red lobes. 
As a result, we identified nine candidates for the blue lobes and four candidates for the red lobes, 
and estimated the lower limits of physical parameters such as the mass, momentum, and energy.
Though some of the identified lobes around the \Hii regions (i.e., B2, B3, R1, and R2)
might not be real outflows but merely tracing the high velocity gas of the expanding shell of the \Hii region,
it is difficult to distinguish the two phenomenon with the present dataset, because of the complex velocity field in W40.

(4) Based on the \hco\4 data, we identified six clumps and derived their physical properties such as the mass,
radial velocities, and surface areas. We further discussed their locations
relative to the \Hii region, and suggested a three-dimensional model of their distributions. 
 

\acknowledgments
We are very grateful to the anonymous referees for their helpful comments and suggestions
to improve this paper. 
We are also grateful to Dr. K. K. Mallick for his kindly providing us with 
the continuum-subtracted 2.12 $\micron$ H$_{2}$ narrowband data.
This work was financially supported by Grant-in-Aid for Scientific Research 
(Nos. 24244017, 26350186, 26610045, and 26287030)
of Japan Society for the Promotion of Science (JSPS). 
The ASTE telescope is operated by National Astronomical Observatory of Japan (NAOJ) 
and the 45 m radio telescope is operated by NRO, a branch of NAOJ. 
\clearpage



\appendix

\section{Derivation of the Column Density and the Fractional Abundance of CCS \label{sec:N(CCS)}}
We estimated the upper limit of $N$(CCS) using Equations (\ref{eq:g}) and (\ref{eq:Nco}) 
in  Appendix \ref{sec:Outflow2} from the noise level of the obtained CCS data
($\sigma \simeq 0.25$ K \kms integrated over the velocity range $3<V_{\rm LSR}<12$ \kms).
We used constants such as $\mu$ and $B_{0}$ in the equations for the CCS molecule 
summarized by \citet[][see their Table 4]{Shimoikura2012},
and used $Q$ of 24.302 for a typical excitation temperature of 
CCS \citep[$T_{\rm ex}=5$ K, e.g,][]{Hirota2009}. To derive $f$(CCS), 
we smoothed the \hco spectrum to the same angular resolution as 
CCS ($57\arcsec$) at the \hco intensity peak position (i.e., Clump 2 in Table \ref{tab:clump}), and derived $N$(H$_2$) from the \hco
spectrum using the $N$(\hco)--$N$(H$_2$) conversion factor reported by \citet{Pirogov}
in the same way described in Section \ref{sec:dense}. We finally estimated $f$(CCS) as $N$(CCS)/$N$(H$_2$).

\section{High Velocity Wings in the $^{12}$CO spectra \label{sec:gaussian_fit}}
\begin{figure*}
\begin{center}
\includegraphics[scale=.8]{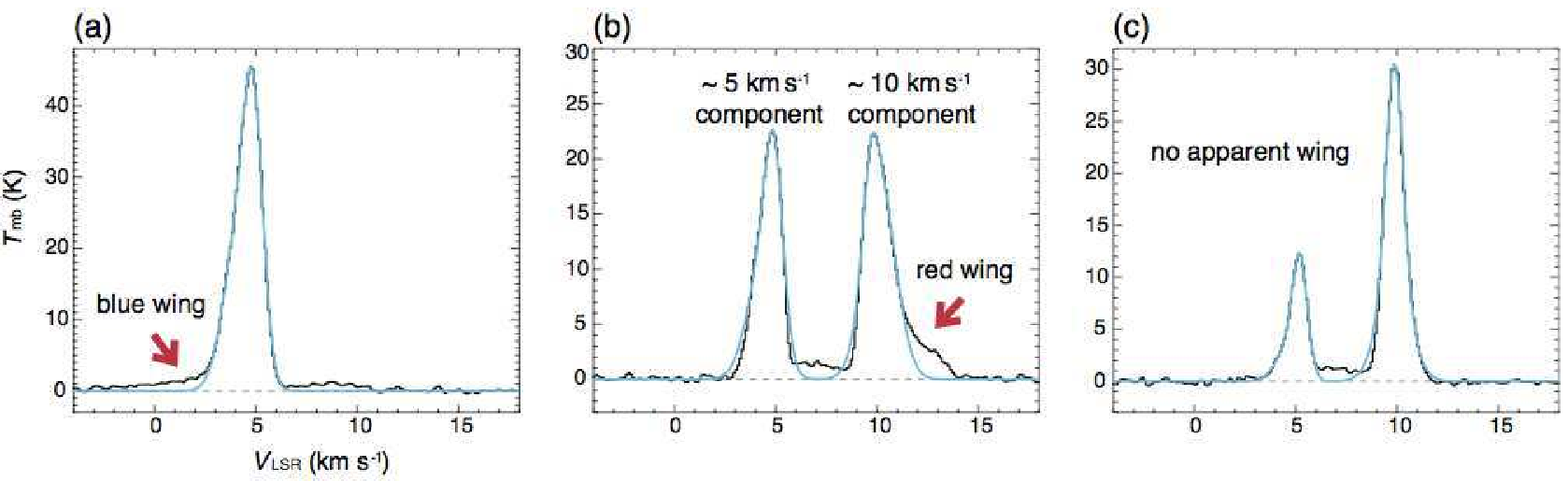}
\caption{
Examples of the Gaussian functions (blue lines) best fitting the observed \co emission lines (black lines)
exhibiting (a) a blue-shifted high velocity wing associated with the $\sim5$ \kms component,
(b) a red-shifted high velocity wing associated with the $\sim10$ \kms component,
and (c) no apparent high velocity wing. Each of the $\sim5$ and  $\sim10$ \kms components
is fitted with two Gaussian functions.
The spectra in panels (a), (b), and (c) are the \co spectra shown in
panels (a), (g), and (h) of Figure \ref{fig:spe}, respectively. 
\label{fig:gaussian}}
\end{center}
\end{figure*}

In order to confirm the existence of the high velocity wings seen in the \co spectra,
we fitted the $\sim5$ \kms and $\sim10$ \kms components by simple Gaussian functions
to see whether the high velocity wings are merely tracing outskirts of the intense \co emission or not.
We first masked the velocity range $V_{\rm LSR }=6-9$ \kms where the \co spectra
are heavily contaminated by the absorption, and then fitted each of the $\sim5$ \kms
and $\sim10$ \kms components with a Gaussian function with two components by
applying a weight proportional to the second power of the observed brightness
temperature at each velocity. We demonstrate some examples of the results in Figure \ref{fig:gaussian}.
Panels (a), (b), and (c) of the figure show the \co spectra exhibiting high velocity wings around the $\sim5$ \kms
component, the $\sim10$ \kms component, and no obvious high velocity wing, respectively,
which are the same spectra shown in Figure \ref{fig:spe}(a), (g), and (h).
As seen in Figure \ref{fig:gaussian}, the \co spectra without apparent high velocity wings
are nicely fitted by the simple Gaussian functions, while those with the apparent wings
cannot be fitted well, indicating that the high velocity wings are not merely the
outskirts of the intense \co emission at $\sim5$ \kms and $\sim10$ \kms.

We should also note that we performed the above Gaussian fitting to all of the \co
spectra in the observed region, and subtracted the fitted lines from the observed spectra
to generate a map of the high velocity wings. The resulting map (not shown) appears
very similar to the blue- and red-shifted lobes shown in Figure \ref{fig:outflow}.


\section{Derivation of Outflows Properties} \label{sec:Outflow2}

In this appendix, 
we explain how we calculated the physical parameters of the outflow lobes. 
We first derived the excitation temperatures $T_ {\rm ex}$ of the observed region 
from the peak brightness temperature of the \co data at each observed position.
In general, the observed brightness temperature $T_ {\rm mb}^{\rm CO}$
can be expressed as
\begin{equation}
\label{eq:radiative}
T_ {\rm mb}^{\rm CO}=[J(T_ {\rm ex})-J(T_{\rm bg})][1-{\rm exp}(-\tau)]
\end{equation}
where  $J(T)=T_{0} /(e^{T_{0} /T}-1)$, $T_{\rm bg}=2.7$ K, and $T_{0}=h\nu_{0} /k$.
Here, $\nu_{0}$ is the rest frequency of the \co\3 emission line, $k$ is the Boltzmann constant,
and $h$ is the Planck constant. The constant $T_{0} $ is therefore 16.59 K for this emission line. 
We derived $T_ {\rm ex}$ from Equation (\ref{eq:radiative}) at each pixel in the \co map
assuming that the line is optically thick ($\tau\gg 1$) at the velocities 
giving the maximum temperature to $T_ {\rm mb}^{\rm CO}$.
The resulting $T_ {\rm ex}$ varies depending on the positions with the maximum value 65.6 K 
and the average value $\sim30$ K. 
We will assume $T_ {\rm ex}=30$ K to derive the physical parameters of the outflow lobes.

In order to estimate the total masses of the outflow lobes, 
we assumed the local thermodynamic equilibrium (LTE). 
The column density of molecule $X$ (e.g., $X=$ \co or \hco),
$N_X$, can be derived from the observed spectra $T_{\rm mb}^{X}(v)$
by the following equation, 
\begin{equation}
\label{eq:NX}
{\rm d}N_{X}= g~T_{\rm mb}^{X}(v)~{\rm d}v
\end{equation} 
where $g$ is a function of $T_ {\rm ex}$, the escape probability $\beta$, and 
some molecular constants, and  is expressed as \citep[e.g.,][]{Hirahara}
\begin{equation}
\label{eq:g}
g=\frac{3h}{8\pi^{3}}\ 
\frac{Q}{\mu^{2}S_{ij}}\ 
\frac{e^{E_{u}/kT_ {\rm ex}}}{e^{T_{0} / T_ {\rm ex}}-1}\ 
\frac{\beta^{-1}}{J(T_ {\rm ex}) - J(T_{\rm bg})}~~~.
\end{equation} 
Here, $Q$ is the partition function and is $Q=k\,T_{\rm ex}/h\,B_{0}$ where
$B_{0}$ is the rotational constant of the molecule. 
$\mu$ is the dipole moment, $E_{u}$ is the energy of the upper level, 
and $S_{ij}$ is the intrinsic line strength of the transition for $i$ to $j$ state. 
For the \co\3 emission line, we used the values of 57.635 GHz, 0.11 debye, and 23.069 cm$^{-1}$
for the constants $B_{0}$, $\mu$, and $E_{u}$, respectively. 
We list these constants in Table \ref{tab:const}.
All of the constants were gathered from JPL Catalog and Splatalogue.
$\beta$ is related to the optical depth $\tau$ as $\beta  = (1 - {e^{ - \tau}})/\tau$.
We assume that the high velocity wings of the \co\3 emission line are optically thin ($\tau\ll 1$), and therefore $\beta=1$.

For these values, Equation (\ref{eq:NX}) for $X=$CO leads to 
\begin{equation}
\label{eq:Nco}
N_{\rm CO}= 4.36\times10^{14}\int_{V_{\rm wing}} T_{\rm mb}^{\rm CO}(v)~{\rm d}v ~~~~{\rm cm^{-2}}
\end{equation} 
where the term $\int T_{\rm mb}^{\rm CO} {\rm d}v$ is in units of K \kms. The velocity range for the integration
is $-3.2\leqq V_{\rm wing}\leqq2.8$ \kms for the blue lobes and 
$11.7\leqq V_{\rm wing}\leqq14.5$ \kms for the red lobes, as stated in Section \ref{sec:Outflow}.

The hydrogen molecular column density $N_{\rm H_2}$ is then calculated 
from $N_{\rm CO}$, assuming the fractional abundance
$N_{\rm CO}/N_{\rm H_2}=1\times10^{-4}$ \citep{Frerking}. 
The mass of the outflow lobes $M_{\rm flow}$ can be obtained 
by integrating the product $\alpha m_{\rm H}N_{\rm H_2}$ 
where $ m_{\rm H}$ is the hydrogen mass and $\alpha$ is the mean molecular weight assumed to be 2.4 
corrected for the $20\%$ helium abundance. 


The integration should be made over the surface area of the outflow lobes $S_{\rm lobe}$ as
\begin{equation}
\label{eq:Mass}
M_{\rm flow}=7.13\times10^{-9} \,D^2 {\int_{S_{\rm lobe}} \int_{V_{\rm wing}}T_{\rm mb}^{\rm CO}}(v){\rm d}v{\rm d} S~~~ M_\sun 
\end{equation} 
where the term ${\int \int T_{\rm mb}^{\rm CO}} {\rm d}v {\rm d}S$ is in units of K \kms arcmin$^2$, and $D$ 
is the distance to the outflow candidates in units of pc ($D=500$ pc for W40). 
We defined $S_{\rm lobe}$ at the 1.2 K \kms (corresponding to the $\simeq9 \sigma$ noise level) contour level as defined in Section \ref{sec:Outflow}. 

\begin{figure*}
\begin{center}
\includegraphics[scale=.8]{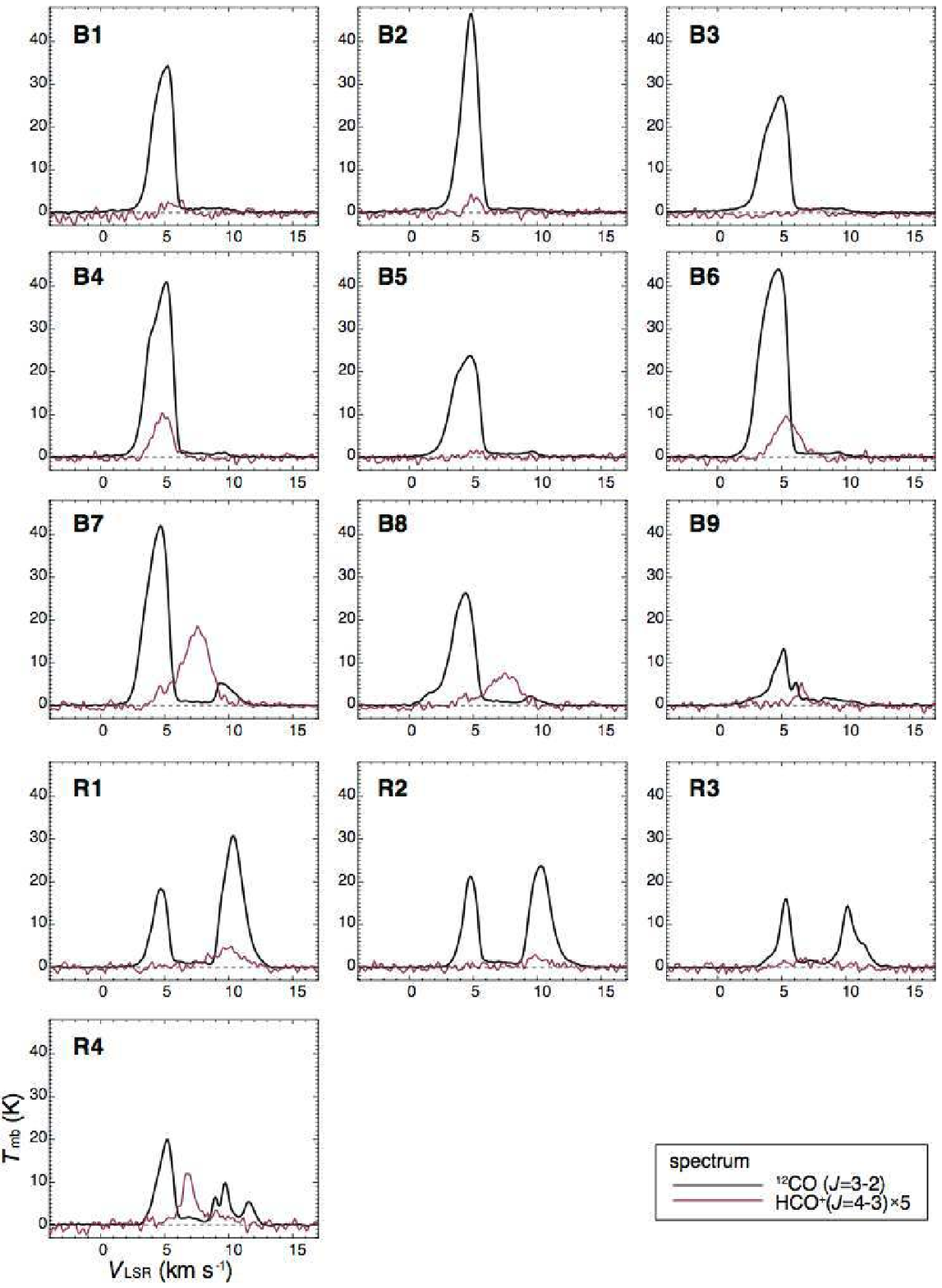}
\caption{
The \co\3 (black) and \hco\4 (red) spectra averaged over the surface areas
of the identified outflow lobes. The \hco spectra are scaled up by a factor of 5.
\label{fig:avg_spe}}
\end{center}
\end{figure*}

The total momentum $P_{\rm flow}$ and kinetic energy $E_{\rm flow}$ contained in the outflow lobes
can be estimated by the following equations,
\begin{equation}
\label{eq:Momentum}
{P_{\rm flow}} = 7.13\times10^{-9} \,D^2 \int_{S_{\rm lobe}} \int_{V_{\rm wing}} {T_{\rm mb}^{\rm CO}}(v)~ {|v-V_{\rm sys}|}~{\rm d}v{\rm d}S
~~~ M_\sun  \,{\rm km \,s^{-1}}
\end{equation} 
\begin{equation}
\label{eq:Energy}
{E_{\rm flow}} = 3.57\times10^{-9} \,D^2 \int_{S_{\rm lobe}} \int_{V_{\rm wing}} {T_{\rm mb}^{\rm CO}}(v)~ {(v-V_{\rm sys})^2} ~{\rm d}v{\rm d} S
~~~ M_\sun  \,{\rm km^{2} \,s^{-2}}
\end{equation} 
where $V{\rm_{sys}}$ is the systemic velocity. 
In order to decide $V\rm_{sys}$, we derived the \hco spectra averaged over the surface areas of each outflow lobe. 
We then fitted the spectra with a single Gaussian function, 
and regarded the peak velocity of the Gaussian function as the systemic velocity $V\rm_{sys}$. 
In Figure \ref{fig:avg_spe}, 
we show the \hco and \co spectra averaged over the surface areas of the identified high velocity lobes.
The \hco emission line of B3 is not detected, for which we assumed $V{\rm_{sys}}=5$ \kms for the lobe.
Using Equations (\ref{eq:Mass})--(\ref{eq:Energy}), we calculated $M_{\rm flow}$, $P_{\rm flow}$, and $E_{\rm flow}$ for the individual outflow lobe. 

Next, we attempted to estimate the dynamical timescale, $t_{\rm d}$, of a outflow lobe. 
In general, $t_{\rm d}$ should be calculated using the maximum separation 
between the outflow lobes and the driving sources. 
For the outflow candidates found in W40, however, we cannot identify 
the definite driving sources and also cannot separate the individual outflow lobes reliably. 
We therefore estimated the dynamical timescale as $t_{\rm d} = R_{\rm flow}/V_{\rm char}$ where
$R_{\rm flow}$ is the radius of the lobes defined as $R_{\rm flow}=\sqrt{S_{\rm lobe}/\pi}$
and $V_{\rm char}$ is the maximum difference between $V_{\rm sys}$ and
the velocities observed in the high velocity wings
detected at $> 3\sigma$ ($\simeq0.5$ K) noise level in the \co spectra observed
at the intensity peak positions of the outflow lobes. 
We summarize the physical parameters of the 13 lobes derived in this appendix in Table \ref{tab:outflow1}.

\color{black}

\clearpage




\begin{thebibliography}{}
\expandafter\ifx\csname natexlab\endcsname\relax\def\natexlab#1{#1}\fi

\bibitem[{{Andr{\'e}} {et~al.}(2010){Andr{\'e}}, {Men'shchikov}, {Bontemps},
  {K{\"o}nyves}, {Motte}, {Schneider}, {Didelon}, {Minier}, {Saraceno},
  {Ward-Thompson}, {di Francesco}, {White}, {Molinari}, {Testi}, {Abergel},
  {Griffin}, {Henning}, {Royer}, {Mer{\'{\i}}n}, {Vavrek}, {Attard},
  {Arzoumanian}, {Wilson}, {Ade}, {Aussel}, {Baluteau}, {Benedettini},
  {Bernard}, {Blommaert}, {Cambr{\'e}sy}, {Cox}, {di Giorgio}, {Hargrave},
  {Hennemann}, {Huang}, {Kirk}, {Krause}, {Launhardt}, {Leeks}, {Le Pennec},
  {Li}, {Martin}, {Maury}, {Olofsson}, {Omont}, {Peretto}, {Pezzuto}, {Prusti},
  {Roussel}, {Russeil}, {Sauvage}, {Sibthorpe}, {Sicilia-Aguilar}, {Spinoglio},
  {Waelkens}, {Woodcraft}, \& {Zavagno}}]{Andre}
{Andr{\'e}}, P., {Men'shchikov}, A., {Bontemps}, S., {et~al.} 2010, \aap, 518,
  L102

\bibitem[{{Arce} {et~al.}(2010){Arce}, {Borkin}, {Goodman}, {Pineda}, \&
  {Halle}}]{Arce}
{Arce}, H.~G., {Borkin}, M.~A., {Goodman}, A.~A., {Pineda}, J.~E., \& {Halle},
  M.~W. 2010, \apj, 715, 1170

\bibitem[{{Bally} {et~al.}(1999){Bally}, {Reipurth}, {Lada}, \&
  {Billawala}}]{Bally1999}
{Bally}, J., {Reipurth}, B., {Lada}, C.~J., \& {Billawala}, Y. 1999, \aj, 117,
  410

\bibitem[{{Bontemps} {et~al.}(2010){Bontemps}, {Andr{\'e}}, {K{\"o}nyves},
  {Men'shchikov}, {Schneider}, {Maury}, {Peretto}, {Arzoumanian}, {Attard},
  {Motte}, {Minier}, {Didelon}, {Saraceno}, {Abergel}, {Baluteau}, {Bernard},
  {Cambr{\'e}sy}, {Cox}, {di Francesco}, {di Giorgo}, {Griffin}, {Hargrave},
  {Huang}, {Kirk}, {Li}, {Martin}, {Mer{\'{\i}}n}, {Molinari}, {Olofsson},
  {Pezzuto}, {Prusti}, {Roussel}, {Russeil}, {Sauvage}, {Sibthorpe},
  {Spinoglio}, {Testi}, {Vavrek}, {Ward-Thompson}, {White}, {Wilson},
  {Woodcraft}, \& {Zavagno}}]{Bontemps}
{Bontemps}, S., {Andr{\'e}}, P., {K{\"o}nyves}, V., {et~al.} 2010, \aap, 518,
  L85

\bibitem[{{Bradshaw} {et~al.}(2015){Bradshaw}, {Offner}, \& {Arce}}]{Bradshaw}
{Bradshaw}, C., {Offner}, S. S.~R., \& {Arce}, H.~G. 2015, \apj, 802, 86

\bibitem[{{Crutcher}(1977)}]{Crutcher}
{Crutcher}, R.~M. 1977, \apj, 216, 308

\bibitem[{{Dobashi}(2011)}]{Dobashi2011}
{Dobashi}, K. 2011, \pasj, 63, 1

\bibitem[{{Dobashi} {et~al.}(1998){Dobashi}, {Sato}, \& {Mizuno}}]{Dobashi1998}
{Dobashi}, K., {Sato}, F., \& {Mizuno}, A. 1998, \pasj, 50, L15

\bibitem[{{Dobashi} {et~al.}(2005){Dobashi}, {Uehara}, {Kandori}, {Sakurai},
  {Kaiden}, {Umemoto}, \& {Sato}}]{Dobashi2005}
{Dobashi}, K., {Uehara}, H., {Kandori}, R., {et~al.} 2005, \pasj, 57, 1

\bibitem[{{Dunham} {et~al.}(2014){Dunham}, {Arce}, {Mardones}, {Lee},
  {Matthews}, {Stutz}, \& {Williams}}]{Dunham}
{Dunham}, M.~M., {Arce}, H.~G., {Mardones}, D., {et~al.} 2014, \apj, 783, 29

\bibitem[{{Dzib} {et~al.}(2010){Dzib}, {Loinard}, {Mioduszewski}, {Boden},
  {Rodr{\'{\i}}guez}, \& {Torres}}]{Dzib}
{Dzib}, S., {Loinard}, L., {Mioduszewski}, A.~J., {et~al.} 2010, \apj, 718, 610

\bibitem[{{Elmegreen} \& {Lada}(1977)}]{Elmegreen}
{Elmegreen}, B.~G., \& {Lada}, C.~J. 1977, \apj, 214, 725

\bibitem[{{Ezawa} {et~al.}(2004){Ezawa}, {Kawabe}, {Kohno}, \&
  {Yamamoto}}]{Ezawa}
{Ezawa}, H., {Kawabe}, R., {Kohno}, K., \& {Yamamoto}, S. 2004, in Society of
  Photo-Optical Instrumentation Engineers (SPIE) Conference Series, Vol. 5489,
  Ground-based Telescopes, ed. J.~M. {Oschmann}, Jr., 763--772

\bibitem[{{Frerking} {et~al.}(1982){Frerking}, {Langer}, \&
  {Wilson}}]{Frerking}
{Frerking}, M.~A., {Langer}, W.~D., \& {Wilson}, R.~W. 1982, \apj, 262, 590

\bibitem[{{Girart} {et~al.}(2000){Girart}, {Estalella}, {Ho}, \&
  {Rudolph}}]{Girart}
{Girart}, J.~M., {Estalella}, R., {Ho}, P.~T.~P., \& {Rudolph}, A.~L. 2000,
  \apj, 539, 763

\bibitem[{{Hirahara} {et~al.}(1992){Hirahara}, {Suzuki}, {Yamamoto},
  {Kawaguchi}, {Kaifu}, {Ohishi}, {Takano}, {Ishikawa}, \& {Masuda}}]{Hirahara}
{Hirahara}, Y., {Suzuki}, H., {Yamamoto}, S., {et~al.} 1992, \apj, 394, 539

\bibitem[{{Hirano} {et~al.}(2006){Hirano}, {Liu}, {Shang}, {Ho}, {Huang},
  {Kuan}, {McCaughrean}, \& {Zhang}}]{Hirano2006}
{Hirano}, N., {Liu}, S.-Y., {Shang}, H., {et~al.} 2006, \apjl, 636, L141

\bibitem[{{Hirota} {et~al.}(2009){Hirota}, {Ohishi}, \&
  {Yamamoto}}]{Hirota2009}
{Hirota}, T., {Ohishi}, M., \& {Yamamoto}, S. 2009, \apj, 699, 585

\bibitem[{{Kirsanova} {et~al.}(2014){Kirsanova}, {Wiebe}, {Sobolev}, {Henkel},
  \& {Tsivilev}}]{Kirsanova}
{Kirsanova}, M.~S., {Wiebe}, D.~S., {Sobolev}, A.~M., {Henkel}, C., \&
  {Tsivilev}, A.~P. 2014, \mnras, 437, 1593

\bibitem[{{Kohno} {et~al.}(2004){Kohno}, {Yamamoto}, {Kawabe}, {Ezawa},
  {Sakamoto}, {Ukita}, {Hasegawa}, {Matsuo}, {Tatematsu}, {Sekimoto}, {Sunada},
  {Saito}, {Iwashita}, {Takahashi}, {Nakanishi}, {Yamaguchi}, {Kamazaki},
  {Sekiguchi}, {Ariyoshi}, {Yokogawa}, {Sugimoto}, {Toba}, {Oka}, {Sakai},
  {Tanaka}, {Takahashi}, {Hayakawa}, {Okuda}, {Muraoka}, {Fukui}, {Onishi},
  {Mizuno}, {Moriguchi}, {Minamidani}, {Mizuno}, {Suzuki}, {Ogawa}, {Yonekura},
  {Asayama}, {Kimura}, {Bronfman}, \& {Aste Team}}]{Kohno}
{Kohno}, K., {Yamamoto}, S., {Kawabe}, R., {et~al.} 2004, in The Dense
  Interstellar Medium in Galaxies, ed. S.~{Pfalzner}, C.~{Kramer},
  C.~{Staubmeier}, \& A.~{Heithausen}, 349

\bibitem[{{K{\"o}nyves} {et~al.}(2010){K{\"o}nyves}, {Andr{\'e}},
  {Men'shchikov}, {Schneider}, {Arzoumanian}, {Bontemps}, {Attard}, {Motte},
  {Didelon}, {Maury}, {Abergel}, {Ali}, {Baluteau}, {Bernard}, {Cambr{\'e}sy},
  {Cox}, {di Francesco}, {di Giorgio}, {Griffin}, {Hargrave}, {Huang}, {Kirk},
  {Li}, {Martin}, {Minier}, {Molinari}, {Olofsson}, {Pezzuto}, {Russeil},
  {Roussel}, {Saraceno}, {Sauvage}, {Sibthorpe}, {Spinoglio}, {Testi},
  {Ward-Thompson}, {White}, {Wilson}, {Woodcraft}, \& {Zavagno}}]{Konyves}
{K{\"o}nyves}, V., {Andr{\'e}}, P., {Men'shchikov}, A., {et~al.} 2010, \aap,
  518, L106

\bibitem[{{Kuhn} {et~al.}(2010){Kuhn}, {Getman}, {Feigelson}, {Reipurth},
  {Rodney}, \& {Garmire}}]{Kuhn}
{Kuhn}, M.~A., {Getman}, K.~V., {Feigelson}, E.~D., {et~al.} 2010, \apj, 725,
  2485

\bibitem[{{Kutner} \& {Ulich}(1981)}]{Kutner}
{Kutner}, M.~L., \& {Ulich}, B.~L. 1981, \apj, 250, 341

\bibitem[{{Mallick} {et~al.}(2013){Mallick}, {Kumar}, {Ojha}, {Bachiller},
  {Samal}, \& {Pirogov}}]{Mallick}
{Mallick}, K.~K., {Kumar}, M.~S.~N., {Ojha}, D.~K., {et~al.} 2013, \apj, 779,
  113

\bibitem[{{Maury} {et~al.}(2011){Maury}, {Andr{\'e}}, {Men'shchikov},
  {K{\"o}nyves}, \& {Bontemps}}]{Maury}
{Maury}, A.~J., {Andr{\'e}}, P., {Men'shchikov}, A., {K{\"o}nyves}, V., \&
  {Bontemps}, S. 2011, \aap, 535, A77

\bibitem[{{Mikami} {et~al.}(1992){Mikami}, {Umemoto}, {Yamamoto}, \&
  {Saito}}]{Mikami}
{Mikami}, H., {Umemoto}, T., {Yamamoto}, S., \& {Saito}, S. 1992, \apjl, 392,
  L87

\bibitem[{{Nakamura} \& {Li}(2007)}]{Nakamura2007}
{Nakamura}, F., \& {Li}, Z.-Y. 2007, \apj, 662, 395

\bibitem[{{Nakamura} {et~al.}(2011){Nakamura}, {Sugitani}, {Shimajiri},
  {Tsukagoshi}, {Higuchi}, {Nishiyama}, {Kawabe}, {Takami}, {Karr},
  {Gutermuth}, \& {Wilson}}]{Nakamura2011}
{Nakamura}, F., {Sugitani}, K., {Shimajiri}, Y., {et~al.} 2011, \apj, 737, 56

\bibitem[{{Nakamura} {et~al.}(2014){Nakamura}, {Sugitani}, {Tanaka},
  {Nishitani}, {Dobashi}, {Shimoikura}, {Shimajiri}, {Kawabe}, {Yonekura},
  {Mizuno}, {Kimura}, {Tokuda}, {Kozu}, {Okada}, {Hasegawa}, {Ogawa}, {Kameno},
  {Shinnaga}, {Momose}, {Nakajima}, {Onishi}, {Maezawa}, {Hirota}, {Takano},
  {Iono}, {Kuno}, \& {Yamamoto}}]{Nakamura2014}
{Nakamura}, F., {Sugitani}, K., {Tanaka}, T., {et~al.} 2014, \apjl, 791, L23

\bibitem[{{Offner} {et~al.}(2011){Offner}, {Lee}, {Goodman}, \&
  {Arce}}]{Offner}
{Offner}, S.~S.~R., {Lee}, E.~J., {Goodman}, A.~A., \& {Arce}, H. 2011, \apj,
  743, 91

\bibitem[{{Pirogov} {et~al.}(2013){Pirogov}, {Ojha}, {Thomasson}, {Wu}, \&
  {Zinchenko}}]{Pirogov}
{Pirogov}, L., {Ojha}, D.~K., {Thomasson}, M., {Wu}, Y.-F., \& {Zinchenko}, I.
  2013, \mnras, 436, 3186

\bibitem[{{Rodney} \& {Reipurth}(2008)}]{Rodney}
{Rodney}, S.~A., \& {Reipurth}, B. 2008, {The W40 Cloud Complex}, ed.
  B.~{Reipurth}, 683

\bibitem[{{Rodr{\'{\i}}guez} {et~al.}(2010){Rodr{\'{\i}}guez}, {Rodney}, \&
  {Reipurth}}]{Rod}
{Rodr{\'{\i}}guez}, L.~F., {Rodney}, S.~A., \& {Reipurth}, B. 2010, \aj, 140,
  968

\bibitem[{{Sawada} {et~al.}(2008){Sawada}, {Ikeda}, {Sunada}, {Kuno},
  {Kamazaki}, {Morita}, {Kurono}, {Koura}, {Abe}, {Kawase}, {Maekawa},
  {Horigome}, \& {Yanagisawa}}]{Sawada}
{Sawada}, T., {Ikeda}, N., {Sunada}, K., {et~al.} 2008, \pasj, 60, 445

\bibitem[{{Schwab}(1984)}]{Schwab}
{Schwab}, F.~R. 1984, in Indirect Imaging. Measurement and Processing for
  Indirect Imaging, ed. J.~A. {Roberts}, 333--346

\bibitem[{{Shimajiri} {et~al.}(2008){Shimajiri}, {Takahashi}, {Takakuwa},
  {Saito}, \& {Kawabe}}]{Shimajiri2008}
{Shimajiri}, Y., {Takahashi}, S., {Takakuwa}, S., {Saito}, M., \& {Kawabe}, R.
  2008, \apj, 683, 255

\bibitem[{{Shimajiri} {et~al.}(2009){Shimajiri}, {Takahashi}, {Takakuwa},
  {Saito}, \& {Kawabe}}]{Shimajiri2009}
---. 2009, \pasj, 61, 1055

\bibitem[{{Shimoikura} \& {Dobashi}(2011)}]{Shimoikura2011}
{Shimoikura}, T., \& {Dobashi}, K. 2011, \apj, 731, 23

\bibitem[{{Shimoikura} {et~al.}(2012){Shimoikura}, {Dobashi}, {Sakurai},
  {Takano}, {Nishiura}, \& {Hirota}}]{Shimoikura2012}
{Shimoikura}, T., {Dobashi}, K., {Sakurai}, T., {et~al.} 2012, \apj, 745, 195

\bibitem[{{Shimoikura} {et~al.}(2013){Shimoikura}, {Dobashi}, {Saito},
  {Matsumoto}, {Nakamura}, {Nishimura}, {Kimura}, {Onishi}, \&
  {Ogawa}}]{Shimoikura2013}
{Shimoikura}, T., {Dobashi}, K., {Saito}, H., {et~al.} 2013, \apj, 768, 72

\bibitem[{{Shuping} {et~al.}(2012){Shuping}, {Vacca}, {Kassis}, \&
  {Yu}}]{Shuping}
{Shuping}, R.~Y., {Vacca}, W.~D., {Kassis}, M., \& {Yu}, K.~C. 2012, \aj, 144,
  116

\bibitem[{{Suzuki} {et~al.}(1992){Suzuki}, {Yamamoto}, {Ohishi}, {Kaifu},
  {Ishikawa}, {Hirahara}, \& {Takano}}]{Suzuki1992}
{Suzuki}, H., {Yamamoto}, S., {Ohishi}, M., {et~al.} 1992, \apj, 392, 551

\bibitem[{{Tokuda} {et~al.}(2013){Tokuda}, {Kozu}, {Kimura}, {Muraoka},
  {Maezawa}, {Onishi}, {Ogawa}, {Nakamura}, {Kuno}, {Takano}, {Iono}, {Kawabe},
  \& {Kameno}}]{Tokuda}
{Tokuda}, K., {Kozu}, M., {Kimura}, K., {et~al.} 2013, in Astronomical Society
  of the Pacific Conference Series, Vol. 476, Astronomical Society of the
  Pacific Conference Series, ed. R.~{Kawabe}, N.~{Kuno}, \& S.~{Yamamoto}, 403

\bibitem[{{Vallee}(1987)}]{Vallee}
{Vallee}, J.~P. 1987, \aap, 178, 237

\bibitem[{{van der Tak} {et~al.}(2007){van der Tak}, {Black}, {Sch{\"o}ier},
  {Jansen}, \& {van Dishoeck}}]{Van2007}
{van der Tak}, F.~F.~S., {Black}, J.~H., {Sch{\"o}ier}, F.~L., {Jansen}, D.~J.,
  \& {van Dishoeck}, E.~F. 2007, \aap, 468, 627

\bibitem[{{Wang} {et~al.}(1994){Wang}, {Jaffe}, {Graf}, \& {Evans}}]{Wang1994}
{Wang}, Y., {Jaffe}, D.~T., {Graf}, U.~U., \& {Evans}, II, N.~J. 1994, \apjs,
  95, 503

\bibitem[{{Wu} {et~al.}(2004){Wu}, {Wei}, {Zhao}, {Shi}, {Yu}, {Qin}, \&
  {Huang}}]{Wu}
{Wu}, Y., {Wei}, Y., {Zhao}, M., {et~al.} 2004, \aap, 426, 503

\bibitem[{{Zeilik} \& {Lada}(1978)}]{Zeilik}
{Zeilik}, II, M., \& {Lada}, C.~J. 1978, \apj, 222, 896

\bibitem[{{Zhu} {et~al.}(2006){Zhu}, {Wu}, \& {Wei}}]{Zhu}
{Zhu}, L., {Wu}, Y.-F., \& {Wei}, Y. 2006, \cjaa, 6, 61

\end{thebibliography}

\clearpage


\begin{deluxetable}{cccl} 
\tablecolumns{2} 
\tabletypesize{\scriptsize}
\tablewidth{0pc} 
\tablecaption{Velocity Components Observed in W40 \label{tab:components}} 
\tablehead{ 
\colhead{$V_{\rm LSR}$} & \colhead{Detection} & \colhead{Detection} & \colhead{Comment}	\\
\colhead{(km s$^{-1}$)} & \colhead{in HCO$^{+}$} & \colhead{in CO} & \colhead{ }	\\
}
\startdata 
$\sim3$	&Yes	&	Yes			&	Faint CO and HCO$^+$ emission which might be unrelated to W40.	\\
$\sim5$ 	&	Yes	&	Yes			&	Intense CO emission ($\gtrsim40$ K) originating from	\\
		&		&				&	the dense gas associated with the expanding shell of the \Hii region. \\
$\sim7$	&	Yes	&	Faint 		&	Systemic velocity component of the entire W40 region \\
		&		&	($\sim 2$ K)	&	causing heavy absorption in the CO emission line. \\
$\sim10$	&	Yes	&	Yes  			&	Intense CO emission ($\gtrsim40$ K) originating from	\\
		&		&				&	the dense gas associated with the expanding shell of the \Hii region.\\
\enddata 
\end{deluxetable} 
\clearpage

\begin{deluxetable}{lccrrrrccccccc} 
\tablecolumns{2} 
\rotate
\tabletypesize{\scriptsize}
\tablewidth{0pc} 
\tablecaption{Properties of the Candidates of Outflows \label{tab:outflow1}}
\tablehead{ 
 \colhead{}   &  \multicolumn{2}{c}{Peak Position} &  \colhead{} &   \colhead{}  &   \colhead{} &   \colhead{}\\
   \cline{2-3} 
  \colhead{Lobe} & \colhead{$\alpha$(J2000)} & \colhead{$\delta$(J2000)} & \colhead{$S_{\rm lobe}$} 
  & \colhead{$\int \int T_{\rm mb}\,dv dS$} &\colhead{$V_{\rm sys}$} & \colhead{$V_{\rm char}$} &\colhead{$M_{\rm flow}$}   
  & \colhead{$P_{\rm flow}$}   & \colhead{$E_{\rm flow}$}  & \colhead{$R_{\rm flow}$} &  \colhead{$t_{\rm d}$} \\
  \colhead{}  & \colhead{($\rm^h\:^m\:^s$)}  &\colhead{(${\arcdeg}\:{\arcmin}\:\:{\arcsec}$)}  & \colhead{(arcmin$^{2}$)}  
  &  \colhead{(K km s$^{-1}$arcmin$^{2}$)} & \colhead{(km s$^{-1}$)}  & \colhead{(km s$^{-1}$)}  
  &\colhead{($M_{\sun}$)}  & \colhead{($M_{\sun}$ km s$^{-1}$)}  & \colhead{($M_{\sun}$ km$^{2}$ s$^{-2}$)} 
  & \colhead{(pc)} & \colhead{(10$^{4}$yr)} 
}
\startdata 
B1	&	18 31 11.0	&	$-2$  02 45	&	0.5 	&	1.0 	&	5.7 	&	5.5 	&	0.002 	&	0.01 	&	0.02 	&	0.06 	&	1.0 	\\
B2	&	18 31 11.7	&	$-2$  03 45	&	1.2 	&	3.8 	&	5.0 	&	6.7 	&	0.007 	&	0.03 	&	0.07 	&	0.09 	&	1.3 	\\
B3	&	18 31 19.0&	$-2$  03  05	&	2.6 	&	12.6 	&	5.0 	&	9.3 	&	0.023 	&	0.08 	&	0.19 	&	0.13 	&	1.4 	\\
B4	&	18 31 11.0	&	$-2$  06 45	&	1.5 	&	4.2 	&	4.8 	&	5.5 	&	0.008 	&	0.03 	&	0.05 	&	0.10 	&	1.8 	\\
B5	&	18 31 11.7	&	$-2$  08 15	&	1.8 	&	7.5 	&	5.2 	&	5.3 	&	0.014 	&	0.05 	&	0.09 	&	0.11 	&	2.0 	\\
B6	&	18 31 15.7&	$-2$  07 25	&	1.8 	&	9.9 	&	5.4 	&	5.0 	&	0.018 	&	0.06 	&	0.11 	&	0.11 	&	2.2 	\\
B7	&	18 31 19.7&	$-2$  06 45	&	1.6 	&	5.8 	&	7.4 	&	6.4 	&	0.011 	&	0.06 	&	0.18 	&	0.11 	&	1.6 	\\
B8	&	18 31 22.3&	$-2$  07 25	&	2.0 	&	14.3 	&	7.3 	&	8.3 	&	0.026 	&	0.14 	&	0.41 	&	0.12 	&	1.4 	\\
B9	&	18 31 53.7&	$-2$  02 25	&	0.7 	&	1.2 	&	6.6 	&	6.1 	&	0.002 	&	0.01 	&	0.03 	&	0.07 	&	1.1 	\\
\tableline
R1	&	18 31 26.3	&	$-2$  05 25	&	1.5 	&	4.9 	&	9.8 	&	4.2 	&	0.009 	&	0.02 	&	0.03 	&	0.10 	&	2.3 	\\
R2	&	18 31 31.7	&	$-2$  05 25	&	1.3 	&	3.8 	&	10.1 	&	4.3 	&	0.007 	&	0.02 	&	0.02 	&	0.10 	&	2.1 	\\
R3	&	18 31 40.3	&	$-2$  02 05	&	0.5 	&	0.8 	&	7.1 	&	6.3 	&	0.002 	&	0.01 	&	0.02 	&	0.06 	&	0.9 	\\
R4	&	18 31 48.3	&	$-2$ 02  05	&	0.3 	&	0.5 	&	6.8 	&	6.6 	&	0.001 	&	0.01 	&	0.01 	&	0.05 	&	0.7 	\\
\enddata 

\tablecomments{ 
The velocity range $V_{\rm wing}$ used in Equations (\ref{eq:Nco})--(\ref{eq:Energy}) to calculate 
$\int T_{\rm mb} dv$ is set to 
$-3.2\leqq V_{\rm LSR}\leqq2.8$ km s$^{-1}$ for the blue lobes (B1--B9), 
and $11.7\leqq V_{\rm LSR}\leqq14.5$ km s$^{-1}$ for the red lobes (R1--R4). 
}

\end{deluxetable} 
\clearpage


\begin{deluxetable}{lccccc} 
\tabletypesize{\scriptsize}
\tablewidth{0pc} 
\tablecaption{Molecular Constants \label{tab:const}} 
\tablehead{ 
  \colhead{Molecule} &\colhead{Transition} & \colhead{$S_{ij}$}    
  & \colhead{$B_{0}$}  & \colhead{$\mu$} &  \colhead{$E_u$}\\
 \colhead{}  &\colhead{}  & \colhead{}  
 & \colhead{(GHz)} & \colhead{(Debye)} & \colhead{(cm$^{-1}$)} 
}
\startdata 

${\rm ^{12}CO}$	&	$J=3-2$	&	3.0 	&	57.6350 	&	0.110 	&	23.0690 	\\
${\rm HCO^{+}}$	&	$J=4-3$	&	4.0 	&	44.5944 	&	3.888 	&	29.7490 	\\
\enddata 

\tablecomments{
All of the constants are taken from JPL Catalog and Splatalogue. 
}

\end{deluxetable} 
\clearpage

\begin{deluxetable}{lccccccrrrc} 
\tabletypesize{\scriptsize}
\setlength{\tabcolsep}{0.02in}
\tablewidth{0pc} 
\tablecaption{Properties of the Clumps \label{tab:clump}} 
\tablehead{ 
 \colhead{}  &  \multicolumn{2}{c}{Peak Position} &  \colhead{} &   \colhead{} &   \colhead{} &   \colhead{}  &   \colhead{} &   \colhead{} &   \colhead{} &   \colhead{}\\
   \cline{2-3} 
  \colhead{clump} & \colhead{$\alpha$(J2000)} & \colhead{$\delta$(J2000)} & \colhead{$T_{\rm mb}$} & \colhead{$V_{\rm LSR}$} & \colhead{$\Delta V$}
  & \colhead{${N}$(HCO$^{+})$}& \colhead{$S_{\rm clump}$} & \colhead{$\int \int T_{\rm mb}\,{\rm d}v {\rm d}S$} & \colhead{$M_{\rm clump}$} & \colhead{Associated}\\
  \colhead{}  & \colhead{($\rm^h\:^m\:^s$)}  &\colhead{(${\arcdeg}\:{\arcmin}\:\:{\arcsec}$)}  & \colhead{(K)} & \colhead{(km s$^{-1}$)}  & \colhead{(km s$^{-1}$)} 
  &  \colhead{(cm$^{-2}$)}  &  \colhead{(arcmin$^{2}$)}  &  \colhead{(K km s$^{-1}$arcmin$^2$)} &  \colhead{($M_{\sun}$)} &  \colhead{Outflow Lobes} 
}
\startdata 
1 	&	18 31 19.0	&	$-2$  07 05	&	3.5 	&	5.7 	&	3.2 	&	5.5$\times10^{12}$	&	8.4 	&	19.3 	&	14.3 	&	B1,B2,B4,B5,B6	\\
	&		&		&		&		&		&		&		&		&		&	(B3)	\\
2 	&	18 31 21.0	&	$-2$  06 25	&	8.1 	&	7.6 	&	2.3 	&	1.1$\times10^{13}$	&	5.3 	&	24.0 	&	17.8 	&	B7,B8	\\
3 	&	18 31 33.7	&	$-2$  04 15	&	4.5 	&	8.0 	&	1.7 	&	3.8$\times10^{12}$	&	3.4 	&	10.2 	&	  7.6 	&	(R3)	\\
4 	&	18 31 46.3	&	$-2$  04 25	&	6.4 	&	6.9 	&	1.5 	&	5.1$\times10^{12}$	&	14.4 	&	35.4 	&	26.2	&	B9,R4	\\
	&		&		&		&		&		&		&		&		&		&	(R3)	\\
5 	&	18 31 26.3	&	$-2$  05 45	&	1.6 	&	10.3 	&	2.3 	&	1.7$\times10^{12}$	&	2.8 	&	4.6 	&	3.4	&	R1,R2	\\
6 	&	18 31 36.3	&	$-2$  03 55	&	3.4 	&	9.1 	&	1.7 	&	2.9$\times10^{12}$	&	1.5 	&	2.7 	&	2.0 	&	\nodata	\\
\enddata
\tablecomments{The mass of Clump No. 6 may be underestimated by $\sim50 \%$, because of the fixed velocity range used to define the clumps
($9 \leqq V_{\rm LSR} \leqq 12$ km s$^{-1}$). The other half of the mass is included in the Clump 3.
The outflow lobe B3 is outside of the surface area of Clump 1, but we regarded them to be associated with each other, because they are close to each other on the sky. 
Another outflow lobe R3 is located right in the middle of Clumps 3 and 4, and we cannot judge which clump R3 is associated with.
}

\end{deluxetable} 
\clearpage


\clearpage

\clearpage

\clearpage


\clearpage

\clearpage

\clearpage

\clearpage

\clearpage

\clearpage

\clearpage

\clearpage


\clearpage

\end{document}